\documentclass[dvips]{elsart}
\usepackage{epsfig}
\usepackage{times}
\usepackage{graphicx}
\usepackage{caption2}
\usepackage{placeins}
\usepackage{rotate}
\begin{document}
\begin{frontmatter}
%
\title{\boldmath Inclusive production of $\rho^{0}(770), f_0(980)$ and
$f_2(1270)$ mesons in $\nu_{\mu}$ charged current interactions.}
%
\centerline{\bf NOMAD Collaboration}
\vskip 0.5cm
\author[Paris]             {P.~Astier}
\author[CERN]              {D.~Autiero}
\author[Saclay]            {A.~Baldisseri}
\author[Padova]            {M.~Baldo-Ceolin}
\author[Paris]             {M.~Banner}
\author[LAPP]              {G.~Bassompierre}
\author[Lausanne]          {K.~Benslama}
\author[Saclay]            {N.~Besson}
\author[CERN,Lausanne]     {I.~Bird}
\author[Johns Hopkins]     {B.~Blumenfeld}
% ...

\address[LAPP]           {LAPP, Annecy, France}                               
\address[Johns Hopkins]  {Johns Hopkins Univ., Baltimore, MD, USA}            
% ...
\address[CERN]           {CERN, Geneva, Switzerland}                          
\address[Lausanne]       {University of Lausanne, Lausanne, Switzerland}      
% ...
\address[Padova]         {Univ. of Padova and INFN, Padova, Italy}            
\address[Paris]          {LPNHE, Univ. of Paris VI and VII, Paris, France}    
% ...
\address[Saclay]         {DAPNIA, CEA Saclay, France}                         
% ...

\thanks[Perugia]         {Now at Univ. of Perugia and INFN, Perugia, Italy}

 \clearpage
\begin{abstract}

The inclusive production of the meson resonances $\rho^{0}(770)$, $f_0(980)$
and $f_2(1270)$ in neutrino-nucleus charged current interactions has been
studied with the NOMAD detector exposed to the wide band neutrino beam
generated by 450 GeV protons at the CERN SPS. For the first time the
$f_{0}(980)$ meson is observed in neutrino interactions.
The statistical significance of its observation is 6 standard deviations.
The presence of $f_{2}(1270)$ in neutrino interactions
is reliably established. The average multiplicity of these three resonances
is measured as a function of several kinematic variables. 
The experimental results are compared to the multiplicities obtained from a
simulation based on the Lund model.
In addition, the average multiplicity of $\rho^{0}(770)$ in
antineutrino - nucleus interactions is measured.
\end{abstract}
\end{frontmatter}

\newpage

\section{Introduction}
\label{sec:introduction}

\hspace*{0.5cm}
Hadron production in high energy neutrino-nucleon charged current (CC)
interactions can be represented as a two-step process.
The first step is the interaction of a W boson with a
quark of the target nucleon changing it into a quark of
a different flavour. The second step is the hadronization
of the quarks during which coloured partons fragment into
colourless hadrons. Perturbative QCD not being applicable
to the soft process of hadronization, phenomenological models
must be used. The most successful models are the string \cite{mod1}
and cluster fragmentation \cite{mod2} models. \\
\hspace*{0.5cm}
A significant fraction of the particles in the hadron jets directly
observed are produced through the production and decay
of a large variety of very short lived particles, the so-called resonances.
Consequently, the study of inclusive meson and baryon resonances is believed 
to reveal more directly the primary interaction mechanism than
the studies of stable particles (such as pions and kaons). Therefore,
resonance production provides a better ground for comparisons between models
and experimental results. Inclusive resonance production has been
extensively studied in $e^{+}e^{-}$ annihilation \cite{e1,e2,e3,e4,e5,e6}, 
photoproduction \cite{g1}, electroproduction by
muons \cite{m1},
in hadronic interactions \cite{h1}. \\  
\hspace*{0.5cm}
The role of the orbitally excited mesons, for example 
$f_0(980)(J^{PC}=0^{++})$ and $f_2(1270)(J^{PC}=2^{++})$,
is of special interest in view of the possibly 
different production mechanisms.
Both mesons have an orbital angular momentum equal to 1 and a
sum of spins of the valence quarks also equal to 1 resulting in
a total spin of 0 for the $f_0(980)$ and of 2 for the $f_2(1270)$. \\
\hspace*{0.5cm}
A number of nonstandard interpretations of the $f_0(980)$ state have been suggested.
One approach regards the $f_{0}(980)$ as a four quark bound state \cite{th1},
another as a two-kaon molecule \cite{th2}. Gribov \cite{th3} has proposed a new
theory of QCD confinement, in which the $f_{0}(980)$ plays the special role 
of a novel vacuum scalar state. Close et al. \cite{th4} noticed that
the signature of Gribov's vacuum scalar state would be an enhanced 
yield in low multiplicity events. Robson \cite{th5} proposed
an interpretation of the $f_{0}(980)$ as a scalar glueball. 
A recent OPAL study of the $f_{0}(980)$ at LEP \cite{e2} does not
confirm the Gribov interpretation. All measured
characteristics of $f_{0}(980)$ production in $Z^{0}$ decays
are consistent with the $f_{0}(980)$ being a conventional
scalar meson. However,
an analysis of the
WA102 data by Close and Kirk \cite{th6} selects the 'enigmatic' $f_{0}(980)$ 
as a glueball candidate together with 
$f_{0}(1500)$, $f_{J}(1710)$ and $f_{2}(1900)$. \\ 
\hspace*{0.5cm}
Bubble chambers used to be the only detectors where inclusive
production of $\rho$ and other resonances in neutrino
interactions could be studied. In such detectors detailed
information about the final state particles in each event can be obtained.
The neutrino experiments \cite{nu5,nu6,nu7,nu8,nu9,nu10,nu11,nu12,nu13,nu14,nu15,nu16,nu17}
studied inclusive production of $\rho^{0}(770)$ and, in some cases,
of $f_{2}(1270)$ resonances.
BEBC WA59 \cite{nu7} studied $\rho(770)$ (all three charge states), 
$\eta(550)$, $\omega(783)$ and $f_{2}(1270)$. The resonance
production properties were measured as a function of the mass of the
hadronic system, of the charged hadron multiplicity and of various
other kinematical variables. \\
\hspace*{0.5cm}
The main problem common to these studies was the small
statistics (the total of all experiments amounts to about 60000 $\nu_{\mu}CC$
events). The level of accuracy corresponding to this statistics
did not necessitate taking into account the reflections of other
resonances (for the definition of reflections see Section~\ref{sec:fit}).
They were not accounted for in most of the previous neutrino experiments.
Due to low statistics there were no attempts
to look for the $f_{0}(980)$ meson in neutrino data. As to the other
orbitally excited meson, BEBC WA21 \cite{nu6}
using 17750 $\nu_{\mu}p$ interactions found 615$\pm$226 $f_{2}(1270)$ mesons. 
But later, in the combined WA21 + WA25 + WA59 + E180 \cite{nu18}
data sample (58866 $\nu_{\mu}CC$ events), no observation of $f_{2}(1270)$
was reported. \\
\hspace*{0.5cm}
All previous experiments reported an overestimation of the average
multiplicities of $\rho^{0}(770)$ by the Lund model by 50-70\%,
while the shapes of the Lund distributions were in agreement
with the experimental distributions. \\
%
%
\hspace*{0.5cm}
 The NOMAD experiment
in 4 years of data taking (1995-1998) has collected
about $1.3 \times 10^6 \nu_{\mu}CC$ interactions.
In addition, a short antineutrino run
has provided about 32000 $\bar{\nu}_{\mu}CC$ events.
The large statistics of the NOMAD data
and the good quality of the event reconstruction make
possible the study of processes that could not be studied in previous
neutrino experiments. \\
\hspace*{0.5cm}
The article is organized as follows:
section~\ref{sec:setup} describes the experimental setup including the
NOMAD detector, the neutrino beam and the NOMAD simulation software;
in section~\ref{sec:vardef} the kinematical variables
used in this analysis are defined;
section~\ref{sec:selection} and section~\ref{sec:fit} describe the
details of the analysis;
the results are presented and discussed in section~\ref{sec:analysis};
section~\ref{sec:systematics} is devoted to the systematic uncertainties;
the multiplicities measured in this analysis are compared with the
results of previous experiments in section~\ref{sec:compprev};
summary and conclusions are given in section~\ref{sec:conclusion}.

\section{The experimental setup}
\label{sec:setup}

\subsection{The NOMAD detector}
\label{subsec:detector}

\begin{figure}[h]
\begin{center}
   \mbox{
     \epsfig{file=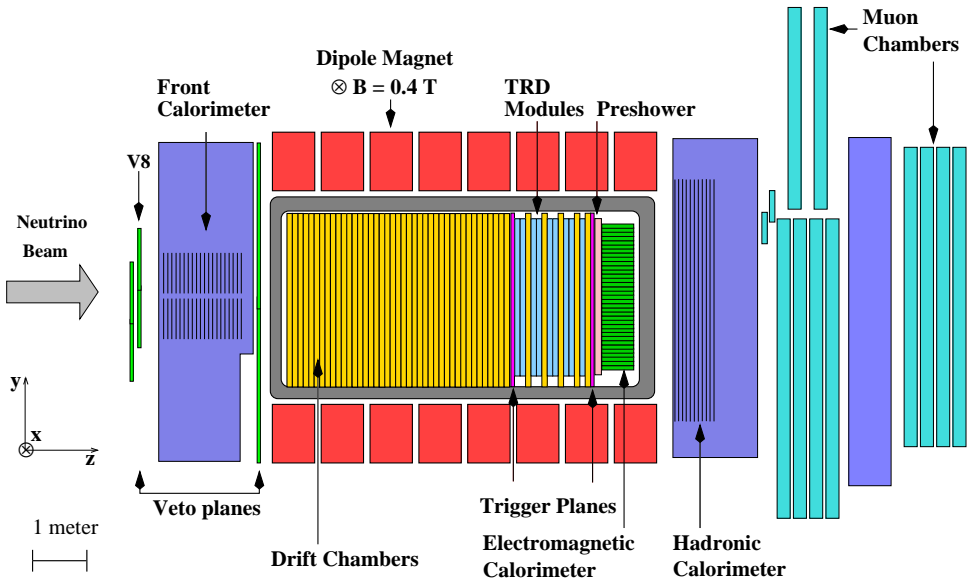,width=130mm}}
     \caption{Side view of the NOMAD detector. The origin of the
coordinate system is on the upstream face of the active target.}
      \label{fig:nomad_detector}
   \end{center}
\end{figure}

\hspace*{0.5cm}
The NOMAD detector (see Fig.~\ref{fig:nomad_detector}) has been described
in details elsewhere \cite{nomad}. 
It consists of a number of sub-detectors most of which are located inside
a 0.4 $T$ dipole magnet with an inner volume of 7.5$\times$3.5$\times$3.5 m$^{3}$:
an active target made of drift chambers (DC) \cite{dc} with a
mass of 2.7 tons, an average density of $0.1$ g/cm$^3$
and a total length of about one radiation length ($\sim 1.0 X_0$),
followed by a transition radiation detector (TRD) \cite{trd}, 
a preshower detector and an electromagnetic
calorimeter (ECAL) consisting of 875 lead glass cells \cite{ecal}.
The low density and the good instrumentation of the active target
(there is less than 1\% of a radiation length between two
consecutive measurements) allow to obtain detailed information
about the final state of neutrino interactions.
The momentum resolution for reconstructed tracks is,
on average, 3.5\% for momenta below 20 GeV/c \cite{nomad}, relevant
to most of the final state hadrons. \\
\hspace*{0.5cm}
The TRD is used for electron identification. It provides a pion
rejection factor greater than $10^3$ for a 90\% electron efficiency in the
momentum range from 1 GeV/c to 50 GeV/c. \\
\hspace*{0.5cm}
The ECAL provides 
an energy resolution of $3.2\%/\sqrt{E \mbox{[$GeV$]} } \oplus 1\%$
for electromagnetic showers and is essential to measure
the total energy flow in neutrino interactions.
The preshower detector is used for electron and gamma identification. \\
\hspace*{0.5cm}
A hadron calorimeter is located just after the magnet coil.
It is followed by an iron absorber and a set of muon chambers.
The muon identification efficiency is 97\% for muons of momentum
higher than 5 GeV/c. \\
\hspace*{0.5cm}
The neutrino interaction trigger \cite{trigger} consists of a coincidence
between signals from two planes of counters located after the active target
in the absence of a signal from a large area system of veto counters
located upstream of the NOMAD detector. \\
\hspace*{0.5cm}
The NOMAD target consists mainly of carbon (on which about 80\%
of the $\nu_{\mu}CC$ interactions occur) and of other elements with
similar atomic numbers (mostly oxygen and nitrogen).

\subsection{The neutrino beam}
\label{subsec:beam}

\hspace*{0.5cm}
The NOMAD detector is located in the WANF neutrino
beam line \cite{wanf} at CERN. In the main focusing mode
(positive focusing) the beam is an essentially pure $\nu_{\mu}$
beam (about 93\%). In negative focusing it is mainly
a $\bar{\nu}_{\mu}$ beam of lower purity (about 71\%).
More details about the beam composition can be found in
Tables ~\ref{tab:beam_info} \cite{beamcalc} and ~\ref{tab:beam_info_negative}.

\begin{table}[htb]
\begin{center}
\begin{tabular}{||c|c|c|c|c||}
\hline
\hline
\multicolumn{1}{||c|}{Neutrino} & \multicolumn{2}{c}{Flux} 
& \multicolumn{2}{|c||}{CC interactions in NOMAD} \\
\cline{2-5}
flavours & $< E_\nu >$ [GeV] & rel.abund. & $< E_\nu >$ [GeV] & rel.abund. \\
\hline
\hline
$\nu_\mu$         & 23.5 & 1    & 43.8 &  1 \\
$\bar{\nu}_\mu$ &  19.2 &  0.0612  &  42.8 &  0.0255 \\
$\nu_e$          &  37.1 &  0.0094  &  58.3 &  0.0148 \\
$\bar{\nu}_e$  & 31.3 &  0.0024 & 54.5 &  0.0016 \\
\hline
\hline
\end{tabular}
\caption{\it The CERN SPS neutrino beam composition in the positive focusing mode.}
\label{tab:beam_info}
\end{center}
\end{table}

\vspace{0.5cm}

\begin{table}[htb]
\begin{center}
\begin{tabular}{||c|c|c|c|c||}
\hline
\hline
\multicolumn{1}{||c|}{Neutrino} & \multicolumn{2}{c}{Flux} 
& \multicolumn{2}{|c||}{CC interactions in NOMAD} \\
\cline{2-5}
flavours & $< E_\nu >$ [GeV] & rel.abund. & $< E_\nu >$ [GeV] & rel.abund. \\
\hline
\hline
$\nu_\mu$         & 23.5 & 0.156 & 58 & 0.383 \\
$\bar{\nu}_\mu$ &  19.6 &  1  & 33.9 & 1 \\
$\nu_e$          &  31.1 &  0.0053  & 56.3 & 0.016 \\
$\bar{\nu}_e$  & 29.5 &  0.0086 & 47.8 & 0.013 \\
\hline
\hline
\end{tabular}
\caption{\it The CERN SPS neutrino beam composition in the negative focusing mode.}
\label{tab:beam_info_negative}
\end{center}
\end{table}


\subsection{The Monte Carlo simulations}
\label{subsec:MC}

\hspace*{0.5cm}
In this analysis the Monte Carlo (MC) simulation software is based on the
LEPTO6.1 \cite{pr1} - JETSET7.4 \cite{pr2} event generator and a full
GEANT321 \cite{pr3} detector simulation. \\
\hspace*{0.5cm}
The simulated events passed through the same reconstruction program
and selection criteria that were used for the data. \\
\hspace*{0.5cm}
 An important JETSET parameter for the fragmentation, the fragmentation
cutoff\footnote{PARJ(33) in JETSET7.4}, was set to 0.2 GeV. The
strangeness suppression factor\footnote{PARJ(2) in JETSET7.4} was
set to 0.21. These values were obtained from a study of hadronic
states \cite{mklepto} with 
invariant masses similar to those studied in this analysis.
This was also checked by comparing the fragmentation function of
pions and $K^0_S$ obtained from the MC with the one obtained
from the NOMAD data. \\
\hspace*{0.5cm}
In order to generate $f_{0}(980)$ and $f_{2}(1270)$ with rates
comparable to those observed in our data we changed the default values
of the following JETSET parameters (the default values for both parameters
in JETSET7.4 is zero): the probability\footnote{PARJ(15) in JETSET7.4}
to produce mesons with orbital angular momentum 1 and total angular momentum 0
equal to 0.08; the probability\footnote{PARJ(17) in JETSET7.4} to produce
mesons with orbital angular momentum 1 and total angular momentum 2
equal to 0.185.  
Note that setting these parameters different from zero decreases the number of
$\rho^0(770)$ mesons in the MC by the sum of values given above
(in our case by 26.5\%). \\
\hspace*{0.5cm}
The MC includes a simulation of nuclear effects \cite{mknucl}
which takes into account the rescattering and absorption of particles
inside nuclei. This allowed us to deconvolute these effects
from our results and extract the resonance yields on free nucleons.
At our energies and target composition these effects are small. \\
\hspace*{0.5cm}
The Monte Carlo allows for a direct test of the fitting
procedure used to extract the number of resonances from the
invariant mass plots since the number of
simulated $\rho^{0}(770)$ mesons is known.
In addition, systematic effects due to other
resonances can be studied since for most of the reconstructed
tracks it is possible to match a MC particle at the level of the event
generator.

\section{Definition of the kinematical variables}
\label{sec:vardef}

\hspace*{0.5cm}
The following variables are used in this analysis:

I. The global event variables:

\begin{itemize}

\item $E_{vis}$ - the total visible energy of the event calculated as
the sum of the energies
of all tracks assigned to the primary vertex and of ECAL clusters
identified as photons. The reconstructed charged particles were assigned
the mass of the pion except for identified muons and electrons which
were assigned their respective masses;

\item $\Delta E_{vis}$ - the measurement error on $E_{vis}$,
calculated from the uncertainty on the fitted momentum for each track and from
the known energy resolution of the ECAL;

\item $P_t^{miss}$ - the modulus of the transverse
component of the total event momentum with respect to the neutrino beam
direction. Lost and mismeasured particles are the main contributors
to the value of $P_t^{miss}$;

\item $E_{\nu}$ - The neutrino energy in the laboratory system.
Monte Carlo studies show that the reconstructed hadronic energy is,
on the average, underestimated, and that this underestimate is
correlated with $P_t^{miss}$. We have therefore corrected
$E_{vis}$ by adding a term depending on $P_t^{miss}$ and
the hadronic energy.
$E_{\nu}$ is determined with a resolution of about 10\% and with
a systematic error not exceeding 1\% in the whole energy range of interest;

\item $\nu = E_{\nu}-E_{\mu}$, where $E_{\mu}$ is the energy of the track
identified as a muon: energy transfer from the neutrino to the hadrons in the
laboratory system. If there were more than one muon candidate, the one
with the greater transverse momentum was chosen;

\item $q = p_{\nu}-p_{\mu}$ - the 4-momentum transfer;

\item $Q^2 = -q^2$ - the 4-momentum transfer squared;

%
\item $W^2 = m_N^2  + 2m_N\nu - Q^2$ - the hadronic invariant mass squared ($m_N$ is the nucleon mass).

\end{itemize}

II. The resonance candidate variables:

\begin{itemize}

\item $x_{F} = 2p_{\parallel}/W$ ($p_{\parallel}$ is the
longitudinal momentum of the resonance candidate in the rest frame
of the hadronic system) - Feynman x variable;

\item $z=E$/$\nu$ (where E is the energy of the resonance candidate) - fragmentation
function variable;

\item $p_{\perp}$ - modulus of the transverse momentum of the resonance candidate
relative to the total momentum of the hadron jet.

\end{itemize}

\section{Data sample and event selection}
\label{sec:selection}

\hspace*{0.5cm}
The present analysis is based on the full NOMAD data sample.
For the $\nu_{\mu}$ analysis we used the data taken with
the neutrino beam set to positive focusing. 
A study of $\bar{\nu}_{\mu}$
in the positive focusing mode would require additional cuts to reduce
the admixture of misidentified $\nu_{\mu}$, which
could render a comparison with other experiments difficult.
For this reason, for the $\bar{\nu}_{\mu}$ study only the negative focusing
mode was used. \\
\hspace*{0.5cm}
A $\nu_{\mu}(\bar{\nu}_{\mu})CC$ event was selected for further processing if
\begin{itemize}

\item there was an identified $\mu^-(\mu^+)$ at the primary vertex with
momentum $p_{\mu}>$3 GeV/c (this cut ensures a good
muon identification and a consistency with previous neutrino
experiments);

\item the invariant mass of the hadronic system $W$ was higher
than 2 GeV (in order to limit the study to the deep inelastic region);

\item there were at least 3 tracks (including the muon) at the
primary vertex (this is necessary to build a resonance candidate).

\end{itemize}

\hspace*{0.5cm}
The following quality cuts were applied in order to ensure an accurate
event reconstruction:

\begin{itemize}

\item the primary vertex (see Fig.~\ref{fig:nomad_detector} for the
coordinate system) should be inside the following fiducial volume:

\begin{center}
         $|X,Y|<120$cm,        $10$cm$<Z<390$cm;    
\end{center}

\item the event should be measured with reasonable accuracy
($\Delta E_{vis}/E_{vis} < 30\%$).

\end{itemize}

After cuts we obtain 669252 $\nu_{\mu}CC$ and 15927 $\bar\nu_{\mu}CC$
event candidates.

\section{Extraction of the resonance signal} 
\label{sec:fit}

\hspace*{0.5cm}
All combinations of tracks with momenta larger than 0.1 GeV/c originating
at the primary vertex and not identified as electrons or muons were used
for the construction of the resonance candidates. All used tracks were
assigned the pion mass.

\begin{figure}[!hb]
\begin{minipage}[t]{0.45\textwidth}
\setcaptionwidth{3cm}
\centering
\includegraphics[height=7.5cm]{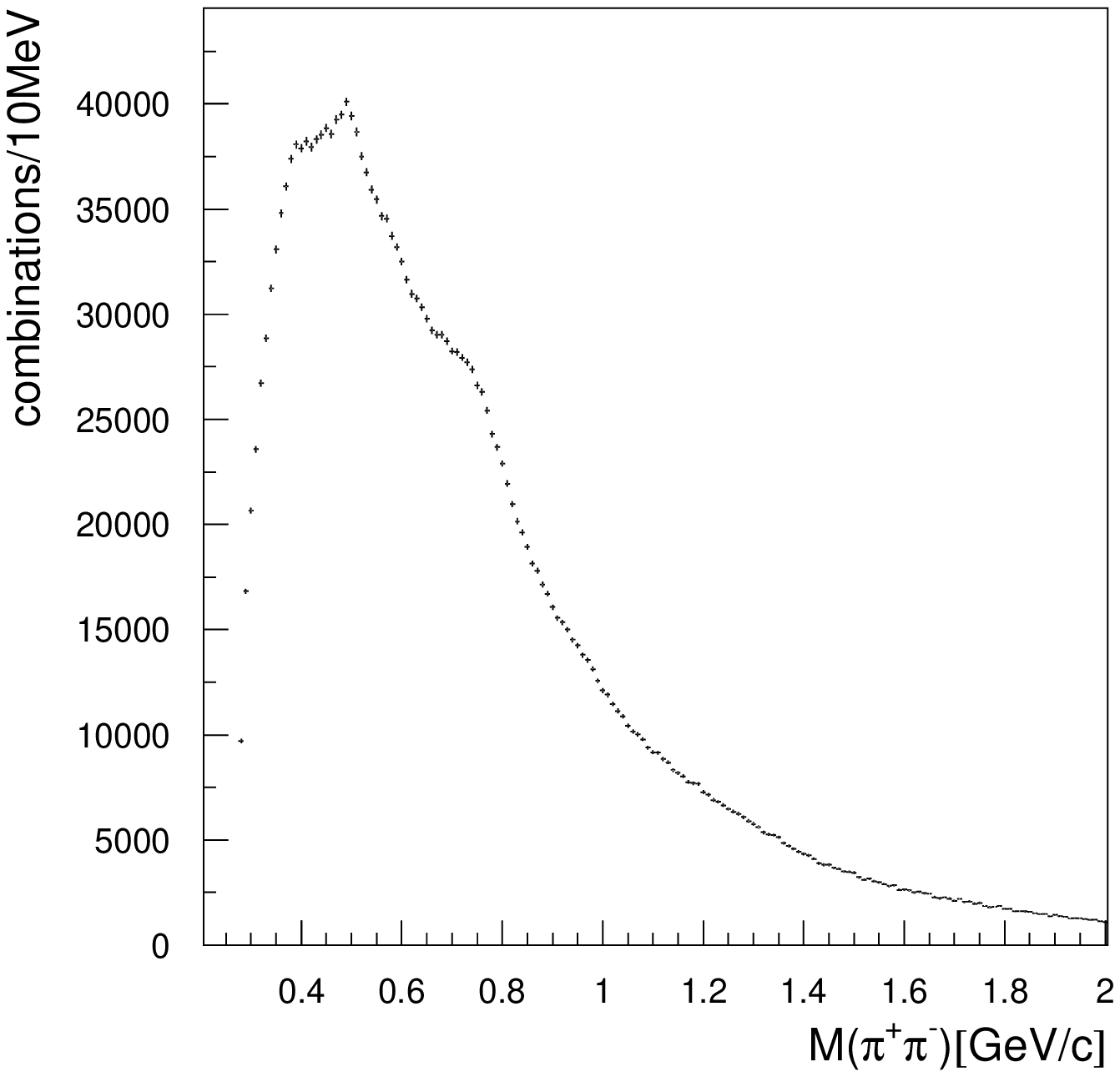}
  \caption{\em Raw distribution of the $\pi^{+}\pi^{-}$ invariant mass
  in the $\nu_{\mu}CC$ sample.}
  \label{figure:r9011}
\end{minipage}
\hspace{1cm}
\begin{minipage}[t]{0.45\textwidth}
\centering
\includegraphics[height=7.5cm]{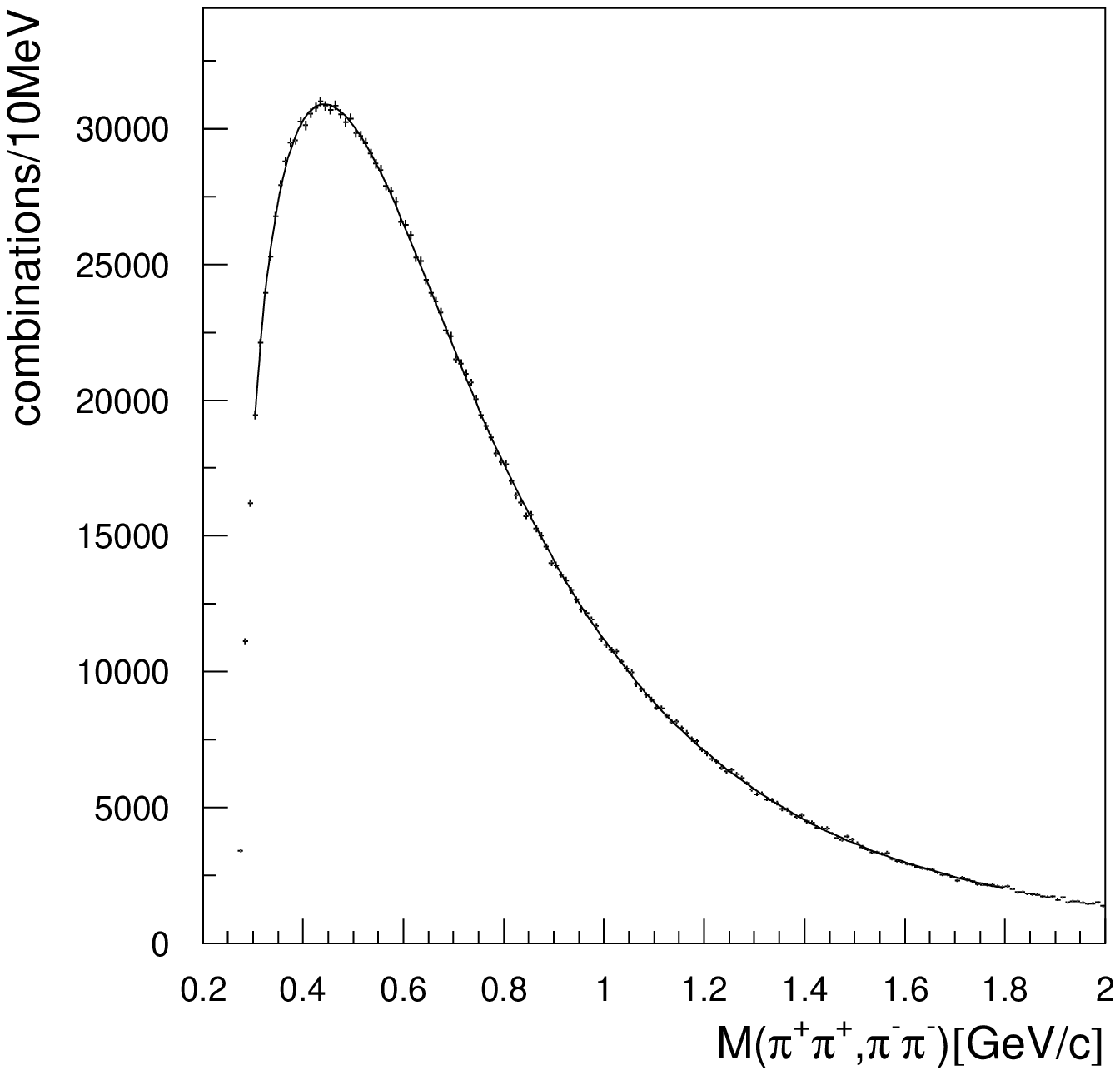}
\caption{\em Raw distribution of the invariant mass of
$\pi^{+}\pi^{+}$ and $\pi^{-}\pi^{-}$ pairs in the $\nu_{\mu}CC$ sample.
The line is the result of the fit described in the text.}
\label{figure:r9101}
\end{minipage}
\end{figure}

\begin{figure}[!hb]
\centering
\includegraphics[height=7.5cm]{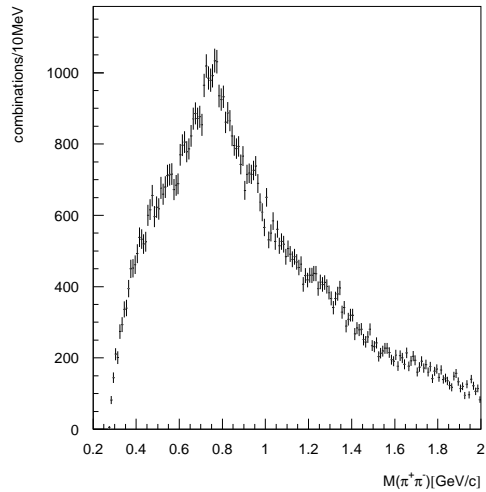}
  \caption{\em Raw distribution of the $\pi^{+}\pi^{-}$ invariant mass
with $x_F>0.6.$}
  \label{figure:xf06}
\end{figure}

\hspace*{0.5cm}
The distribution of the $\pi^{+}\pi^{-}$ invariant mass
is shown in Fig.~\ref{figure:r9011}. 
For comparison the distribution of the invariant mass of
$\pi^{+}\pi^{+}$ and $\pi^{-}\pi^{-}$ pairs
is shown in Fig.~\ref{figure:r9101}. 
The $\pi^{+}\pi^{-}$ distribution shows an enhancement
at the $\rho^{0}(770)$ mass which is not present in the like-sign distribution.
When a cut $x_F > 0.6$ is applied to reduce the combinatorial background,
clear peaks at the $f_{0}(980)$ and $f_{2}(1270)$ masses become
visible, as shown in Fig.~\ref{figure:xf06}. \\
\begin{figure}[t]
\begin{center}
\setcaptionwidth{9cm}
\includegraphics[width=15cm]{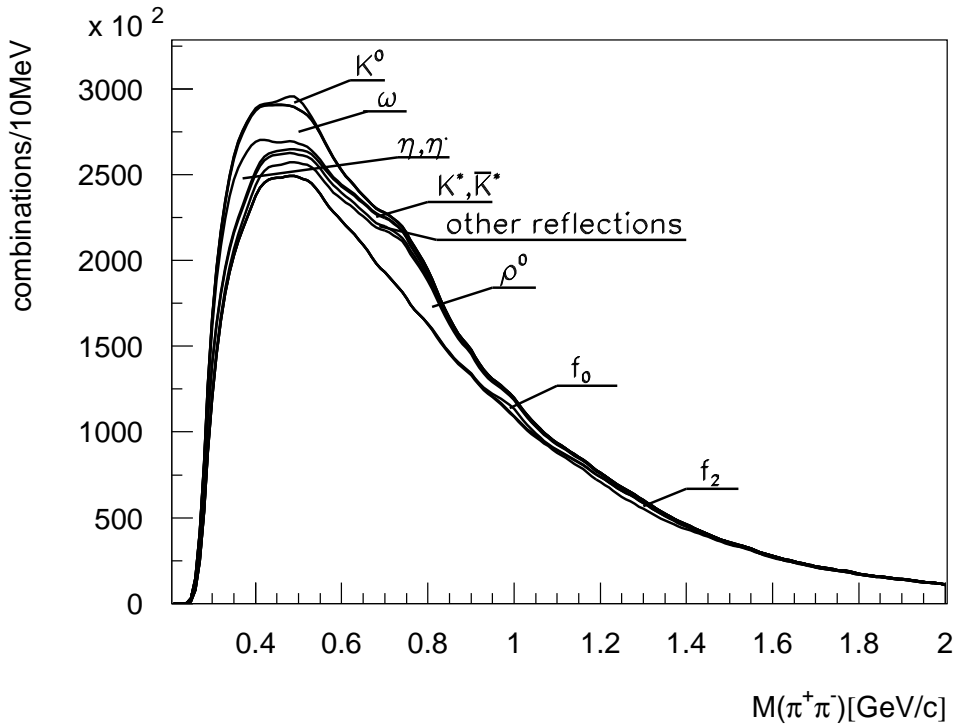}
\caption{\em $\pi^{+}\pi^{-}$ invariant mass distribution of Monte Carlo events
including the contributions of signal, reflections and combinatorial
background.}
\label{figure:reflsub}
\end{center}
\end{figure}
\hspace*{0.5cm}
We now address the treatment of reflections.
Reflections are contributions to the $\pi^+\pi^-$
invariant mass distribution from resonances decaying to $\pi^+\pi^-$ + some
other particles (e.g. $\omega \rightarrow \pi^+\pi^-\pi^0$)
or contributions from resonances decaying to particles other than pions
and thus assigned wrong mass values.
The reflections were estimated from the fully simulated MC events
and taken into account in the fit of the $\pi^+\pi^-$ invariant
mass distribution (Fig.~\ref{figure:reflsub}). It is possible however that
the multiplicity of the resonances giving reflections in the data could
be different from the one in the MC. In order to take this into account
the sum of all reflections obtained from the MC was multiplied by an
overall normalization factor $p$ which was left as a free parameter
in the fit.
In NOMAD an additional contribution to the $\pi^+\pi^-$ invariant mass
distribution arises from early decays of $K^0_S$ and misidentified $\gamma$
conversions (see Fig.~\ref{figure:reflsub}). For distances less than 15cm
from the primary vertex along the beam axis it is difficult
to distinguish tracks beginning at the
primary vertex from tracks originating at a $V^0$ vertex.
This contribution was also estimated from the MC simulation and taken
into account together with other reflections. As can be seen
in Fig.~\ref{figure:reflsub}, the largest contribution is from
$\omega \rightarrow \pi^+\pi^-\pi^0$. \\
\hspace*{0.5cm}
Ignoring reflections could result in incorrect determination of
the resonance parameters. As an example, in our case this would result
in a $\rho^{0}$ mass value extracted from the fit about 10 MeV lower
than the accepted value \cite{bw2}. \\
\hspace*{0.5cm}
 The resonance signal
is determined by fitting the invariant mass distribution of all
possible $\pi^{+}\pi^{-}$ combinations, $dN/dm$, to the expression 
\begin{equation}
   \frac{dN}{dm} = [1+BW(m)]BG(m) + p\cdot RF(m),
\end{equation}
where
\begin{equation}
 BW(m) = a_{1}BW_{\rho}(m)+a_{2}BW_{f_0}(m)+a_{3}BW_{f_2}(m)
\end{equation}
is a relativistic Breit-Wigner function \cite{bw1},
BG(m) is the combinatorial background
and $RF(m)$ is the contribution from reflections. \\
\hspace*{0.5cm}
The Breit-Wigner function is
\begin{equation}
     BW=\frac{m}{k}\frac{m_R\Gamma'_R} {(m^{2}-m^{2}_R)^{2} +
m^{2}_R\Gamma'^{2}_R} 
\end{equation}
where
\begin{equation}
  \Gamma'_R=\Gamma_R(\frac{k} {k_R})^{2L+1}
\frac{m_R}{m}
\end{equation}
\hspace*{0.5cm}
Here $m_R$ and $\Gamma_R$ are the mass and width
of the resonance R; L is the relative orbital angular momentum of the two pions
(equal to the spin of the resonance): L=0 for $f_{0}(980)$,
L=1 for  $\rho^{0}(770)$, and L=2 for $f_{2}(1270)$;
$k$ is the pion momentum in the resonance rest frame;
$k_R$ is the value of $k$ when $m=m_R$. \\
\hspace*{0.5cm}
The background was assumed to have the following shape
\begin{equation}
 BG=a_{4} (m-2m_{\pi})^{a_{5}}
exp(a_{6}m+a_{7}m^{2}+
a_{8}m^{3})
\end{equation}
which takes into account the threshold effect and the exponential fall-off
of the distribution at high values of $m$. \\
\hspace*{0.5cm}
In Equation (1) the Breit-Wigner functions are multiplied by BG
in order to account for the available phase-space.
The parameters $a_{1}$ to $a_{8}$, $p$ and
the masses and widths of the three resonances are free fit parameters.
The mass range covered by the fit was 0.3 GeV$<m<$2 GeV. \\
\hspace*{0.5cm}
As a cross-check of the reliability of the procedure we
performed the fit using another parameterization suggested
by A.Minaenko \cite{Minaenko}.
The $\rho^{0}(770)$ yield found by this method agrees with our
result within 0.5\%. \\
\hspace*{0.5cm}
 A fit to the distribution of the invariant mass for like sign pairs
(Fig.~\ref{figure:r9101}) gives a resonance yield at $m_R = 770 MeV$
compatible with zero within one standard deviation. \\
\hspace*{0.5cm}
In order to take into account the experimental resolution
on the $\pi^{+}\pi^{-}$ invariant mass we followed the
method described in \cite{g1}. We estimated the invariant mass
resolution function from the errors on the track parameters in
the data as given by the reconstruction program.
This function has a FWHM of about 25 MeV and it 
was convoluted with expression (1) for the fit.\\
\hspace*{0.5cm}
The procedure used to obtain the results presented in
section~\ref{sec:analysis} proceeded through the steps listed below.
Steps (i) and (ii) refer to the procedure of obtaining the overall
average multiplicities and the resonance widths and masses, while
the subsequent steps were made to obtain the differential distributions
for the kinematical variables.

\begin{enumerate}

\item Expression (1) is fitted to the raw inclusive experimental
distribution.

\item The "fully simulated" MC is used to estimate the reconstruction
efficiency of various resonances by comparing the yields of reconstructed
and generated resonances. The uncorrected yields measured in the data
are then divided by these efficiencies.
\item A fit to the experimental distributions in each bin of kinematical
variables is performed.
The reflection normalization factor $p$ and the masses and widths
of the resonances are fixed at the values found in step (i).

\item Correction factors are obtained in each of the kinematical
variable bins similarly to step (ii). 
The yields of the resonances in each bin are corrected accordingly.

\item The corrected experimental distributions are plotted and
compared with the MC distributions at the event generator level. \\

\end{enumerate}
\hspace*{0.5cm}
The $\bar\nu_{\mu}CC$ sample was treated in the same way, however,
due to its lower statistics we limited the analysis to the steps (i) - (ii)
and to the $\rho^{0}(770)$ resonance only.

\section{The analysis} 
\label{sec:analysis}

\subsection{Resonance multiplicities and properties}

\begin{figure}[!hb]
\setcaptionwidth{7cm}
\begin{minipage}[t]{0.45\linewidth}
\centering
\includegraphics[height=7.5cm]{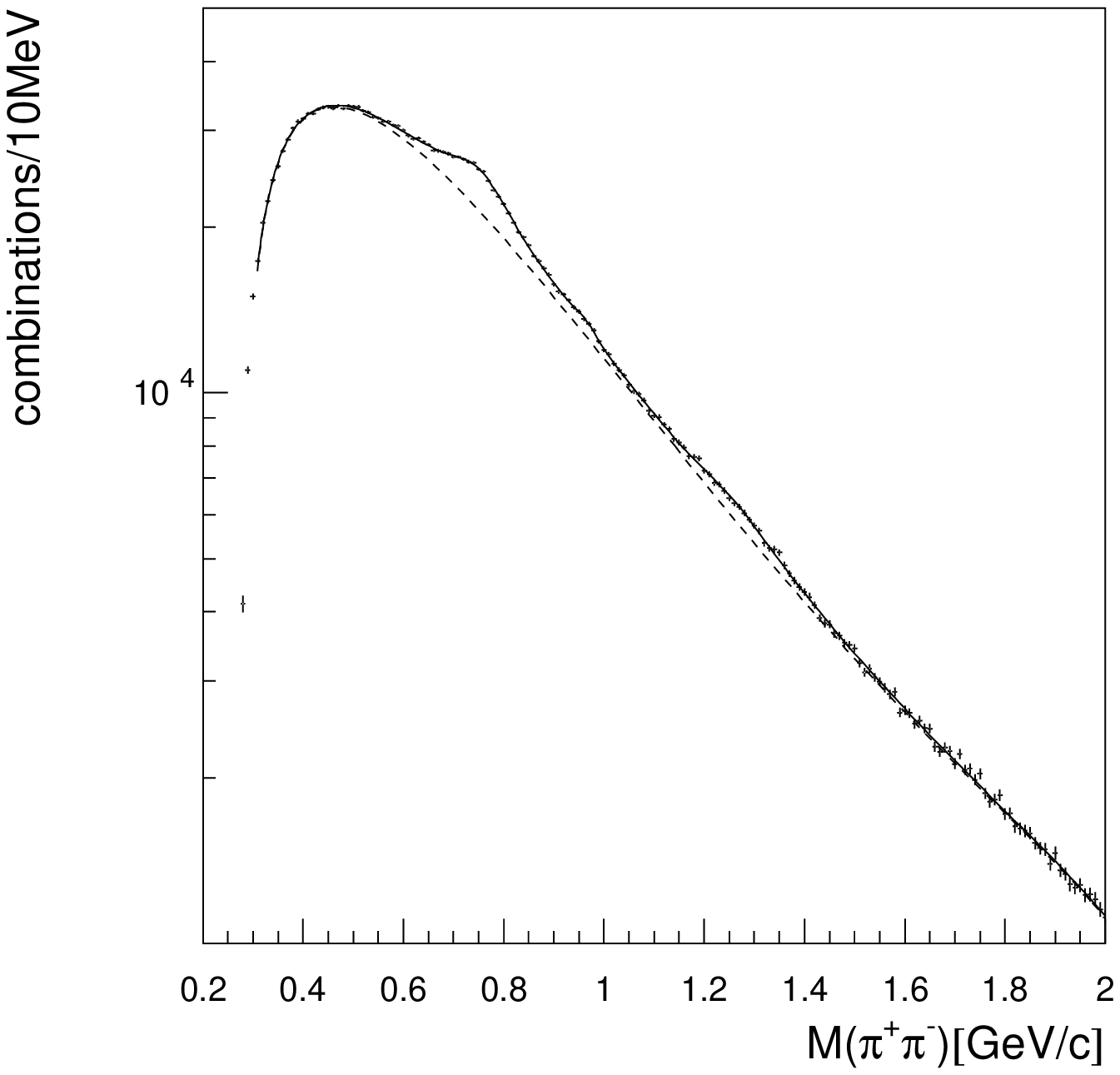}
\caption{\em Distribution of the $\pi^{+}\pi^{-}$ invariant mass
after subtracting the contribution of reflections.}
\label{figure:rhomfit}
\end{minipage}
\hspace{1cm}
\begin{minipage}[t]{0.45\linewidth}
\centering
\includegraphics[height=7.5cm,width=7.5cm]{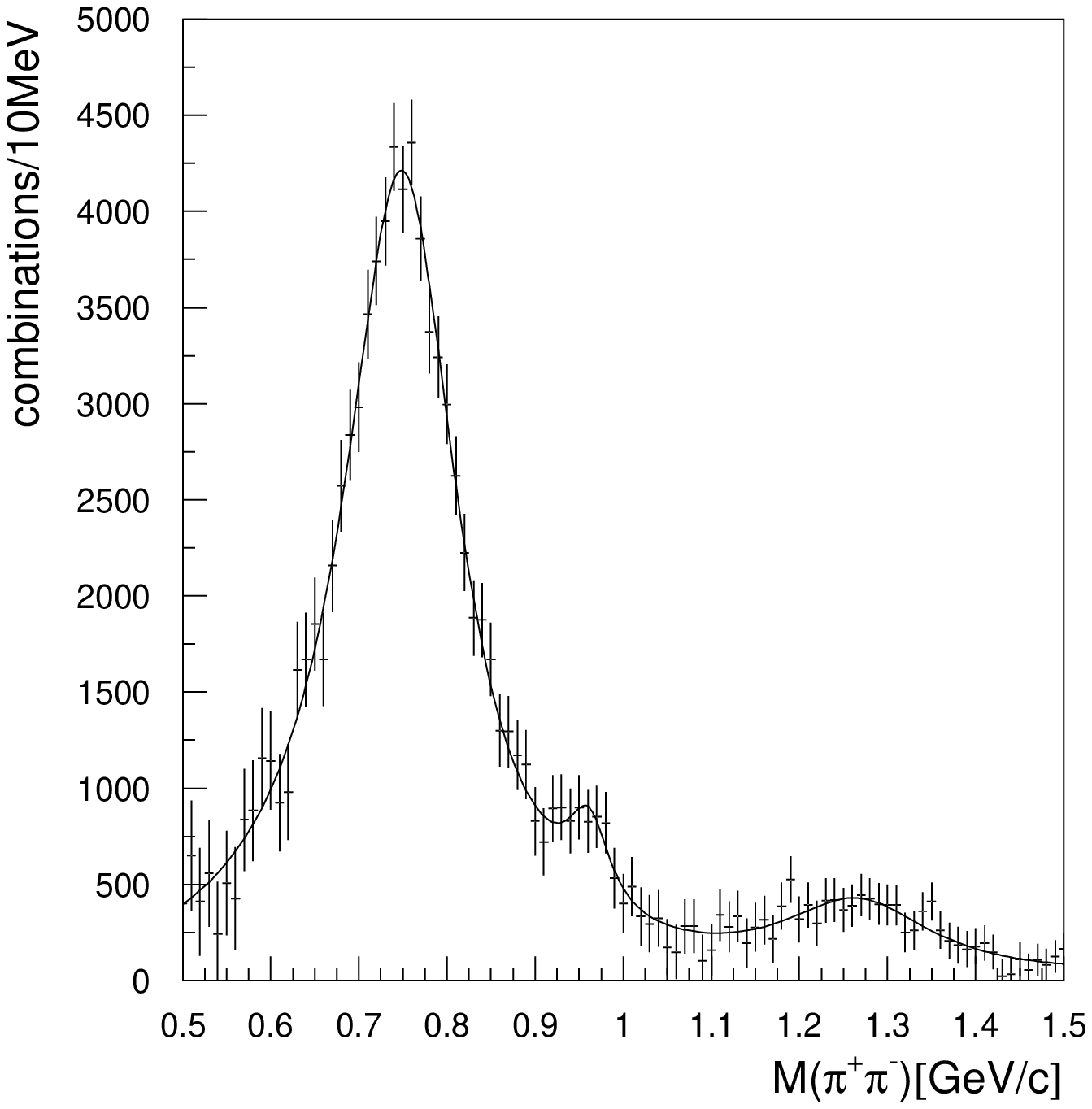}
  \caption{\em Distribution of the $\pi^{+}\pi^{-}$ invariant mass
after subtracting the contributions of both reflections and
combinatorial background.}
  \label{figure:rff-k0}
\end{minipage}%
\end{figure}

\hspace*{0.5cm}
 All the results reported in this section are obtained from
the analysis of the $\nu_{\mu}CC$ sample. \\
\hspace*{0.5cm}
In Fig.~\ref{figure:rhomfit} we show the data after subtracting
the contribution of the reflections as obtained from the fit.
In Fig.~\ref{figure:rff-k0} we further subtract the contribution
of the combinatorial background. The three resonances are 
clearly visible. \\
\hspace*{0.5cm}
After correcting the result of the fit for the efficiencies
as determined from the MC (about 0.77 for all the studied resonances)
and for the branching ratios $BR(R \rightarrow  \pi^+\pi^-)$ \cite{bw2}
we obtain the results shown in Table~\ref{table:resnumb}.

\begin{table}[hbt]
\setcaptionwidth{10cm}
\begin{center}
\begin{tabular}{|l|c|c|c|c|c|} 
\hline
 Resonance      & Branching   &  Number of    & Average      &  Mass &    $ \Gamma$(MeV)\\
     & ratio $\pi^+\pi^-$ \cite{bw2} & resonances   &   Multiplicity  & (MeV) &        \\ 
\hline    
 $\rho^{0}(770)$ & 1.000 & 130368$\pm$4336 & 0.195$\pm$0.007 & 768$\pm$2 &151$\pm$7\\
\hline
 $f_{0}(980)$ & 0.666& 11809$\pm$1965 & 0.018$\pm$0.003 & 963$\pm$5 & 35$\pm$10\\      
\hline
 $f_{2}(1270)$& 0.564 & 25189$\pm$3958 & 0.038$\pm$0.006 & 1286$\pm$9 &198$\pm$30\\
\hline
\end{tabular}
\end{center}
\caption{\em Corrected numbers and multiplicities of the three resonances
and their measured masses and widths. The errors are statistical only.}
\label{table:resnumb}
\end{table}  

\FloatBarrier

\vspace{0.5cm}
\subsection{Resonance multiplicities as a function of the kinematical variables} 
\label{subsec:resdistr}

\hspace*{0.5cm}
These results were obtained as described in the section~\ref{sec:fit},
steps (iii) - (v). For all bins of all kinematical distributions
the resonance signals
were clearly visible and the value of the reduced $\chi^2$ 
was around 1.0. All the numbers corrected for experimental efficiencies
and branching ratios, are compared with the results of
the modified Lund simulation (see section~\ref{subsec:MC}).
The errors shown in the figures are statistical only. \\
\begin{figure}[!hb]
\begin{minipage}[t]{0.33\linewidth}
\centering
\includegraphics[height=5.2cm]{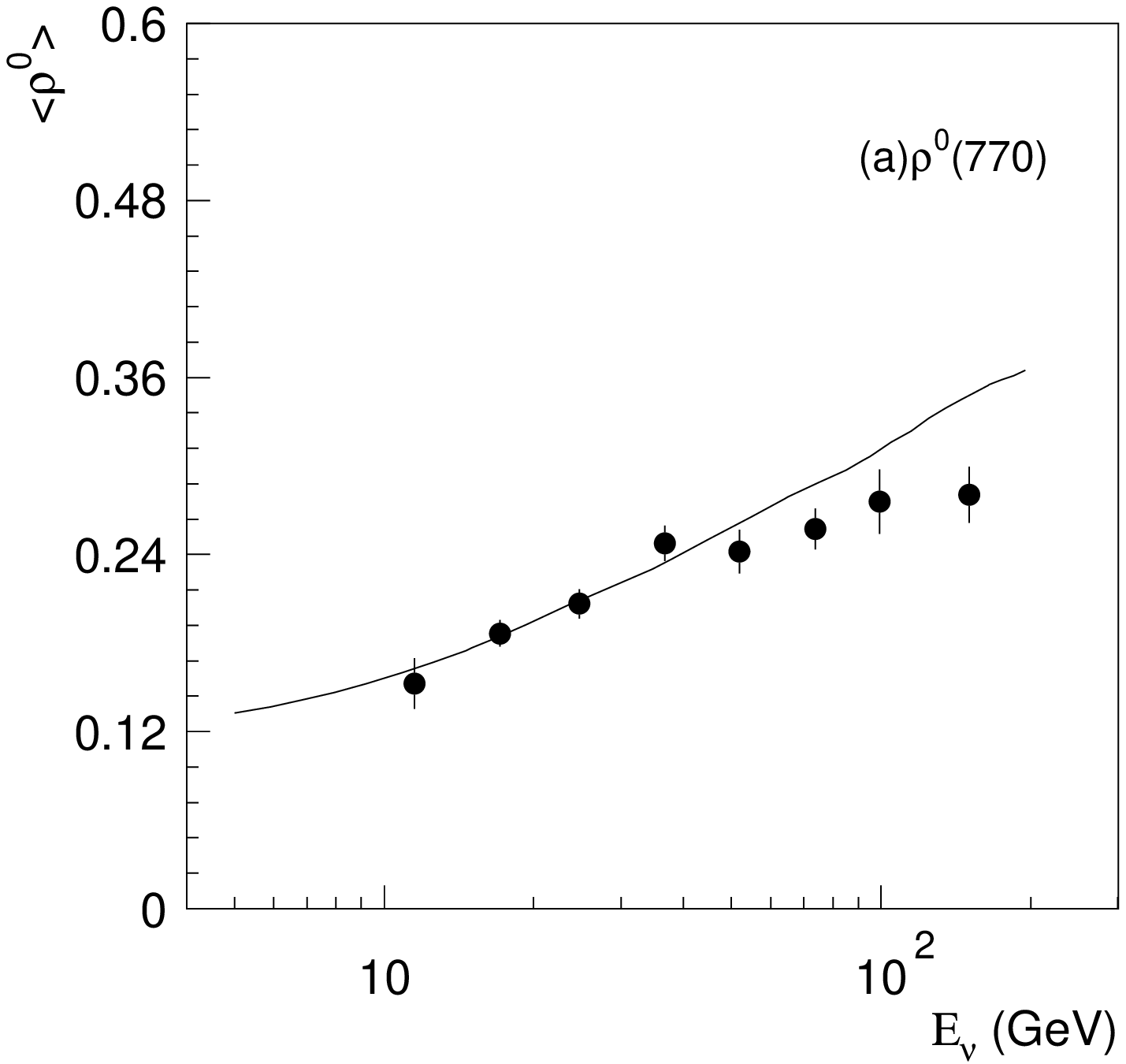}
\end{minipage}%
\begin{minipage}[t]{0.33\linewidth}
\centering
\includegraphics[height=5.2cm]{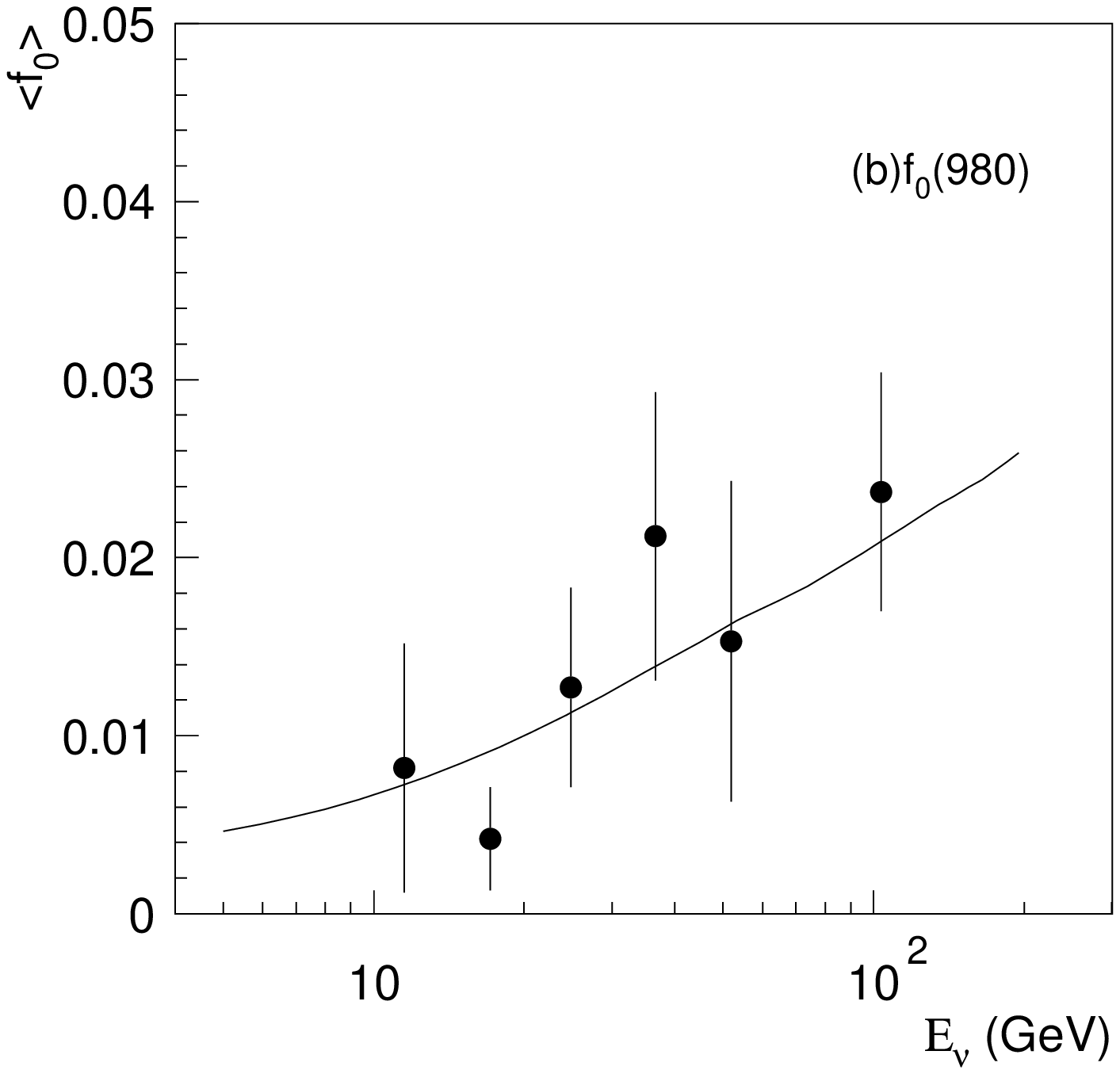}
\end{minipage}%
\begin{minipage}[t]{0.33\linewidth}
\centering
\includegraphics[height=5.2cm]{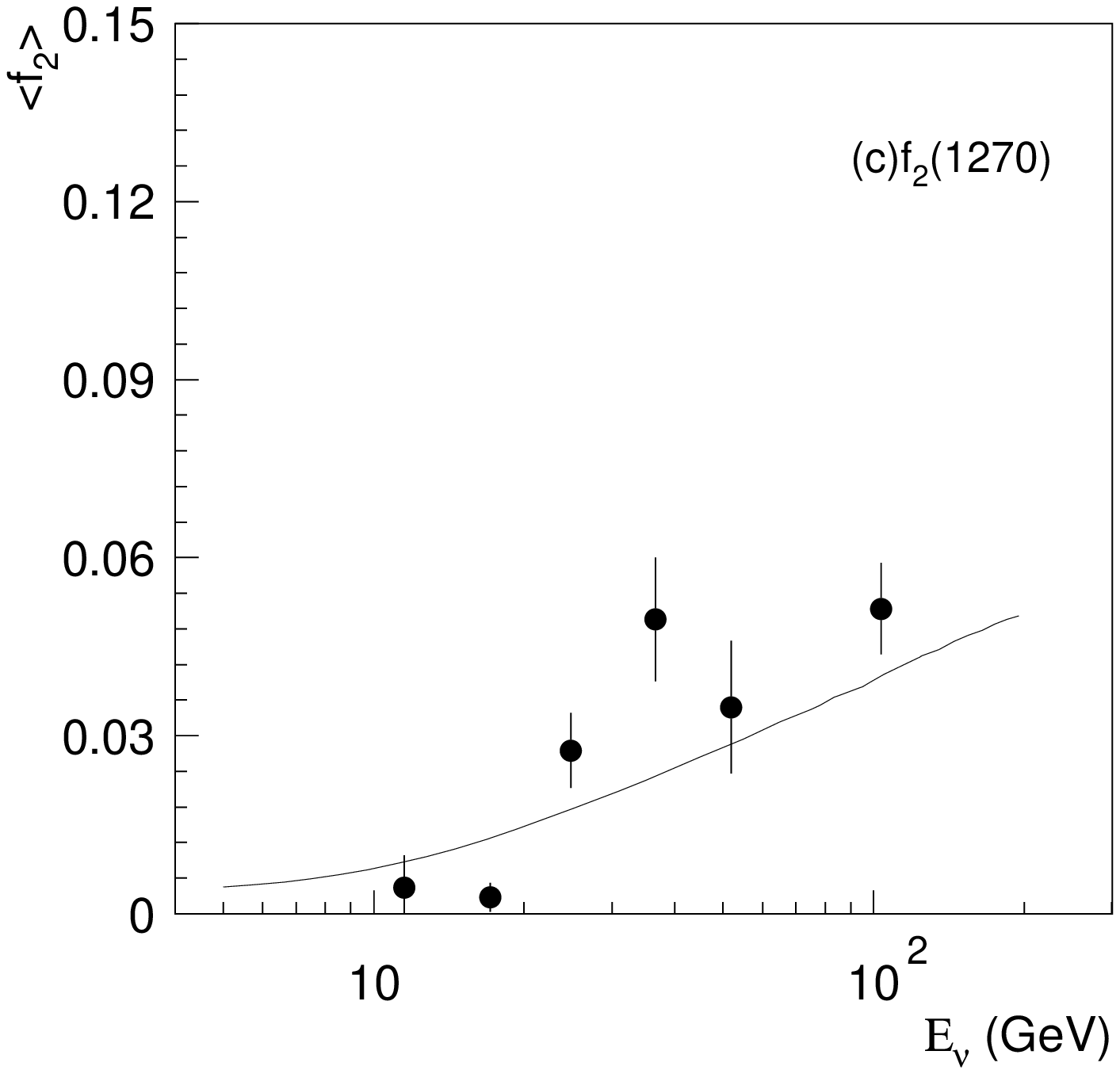}
\end{minipage}%
\caption{\em Average $\rho^{0}(770)$(a), $f_{0}(980)$(b) and $f_{2}(1270)$(c)
multiplicity as a function of $E_{\nu}$.
The solid line represents the result of the modified Lund simulation.
The errors are statistical only.}
\label{figure:enu}
\end{figure}
\hspace*{0.5cm}
The increase of the average multiplicity of the three resonances
as a function of $E_{\nu}$ is shown in 
Fig.~\ref{figure:enu}.

\begin{figure}[!hb]
\begin{minipage}[t]{0.33\linewidth}
\centering
\includegraphics[height=5.2cm]{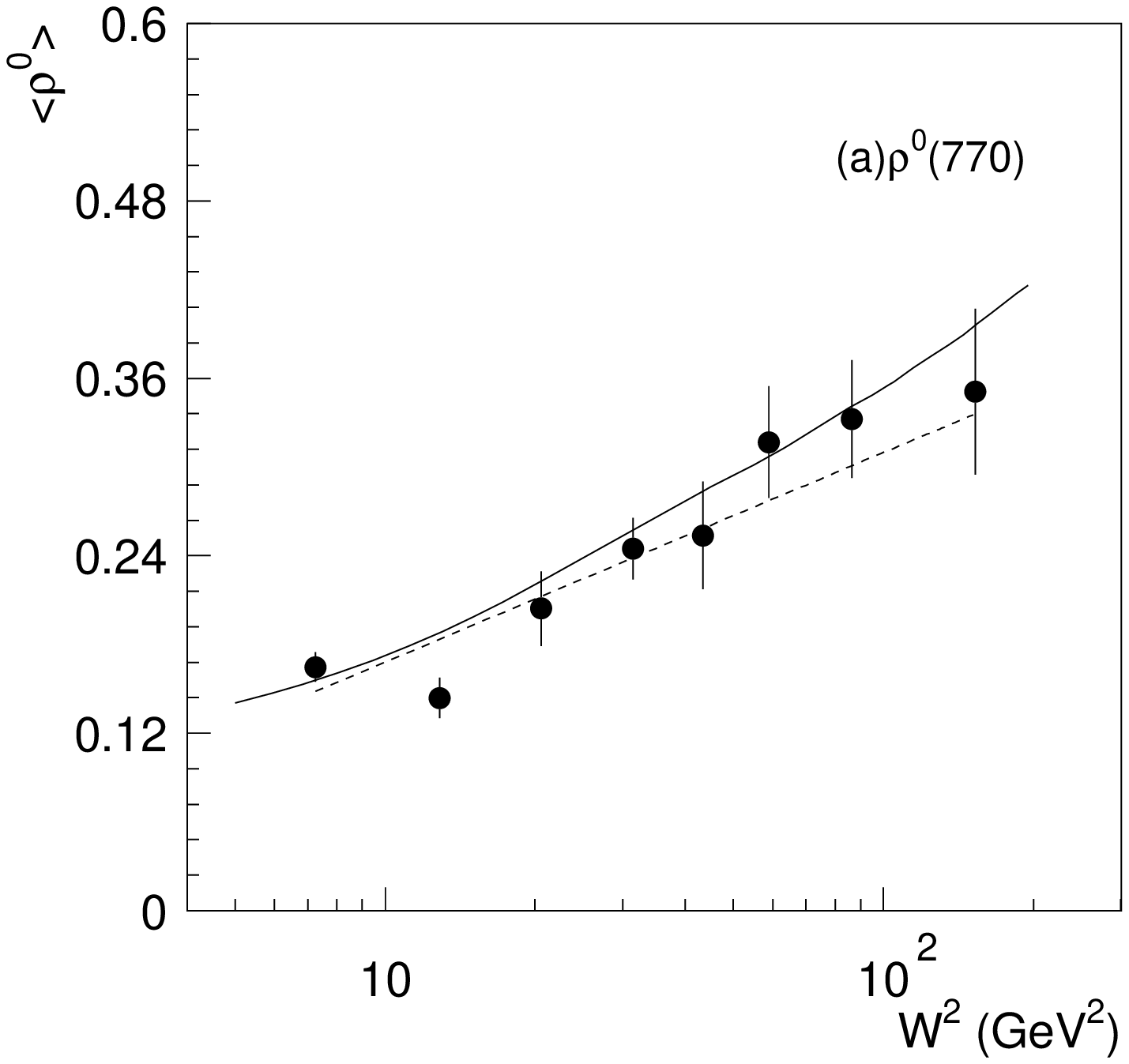}
\end{minipage}%
\begin{minipage}[t]{0.33\linewidth}
\centering
\includegraphics[height=5.2cm]{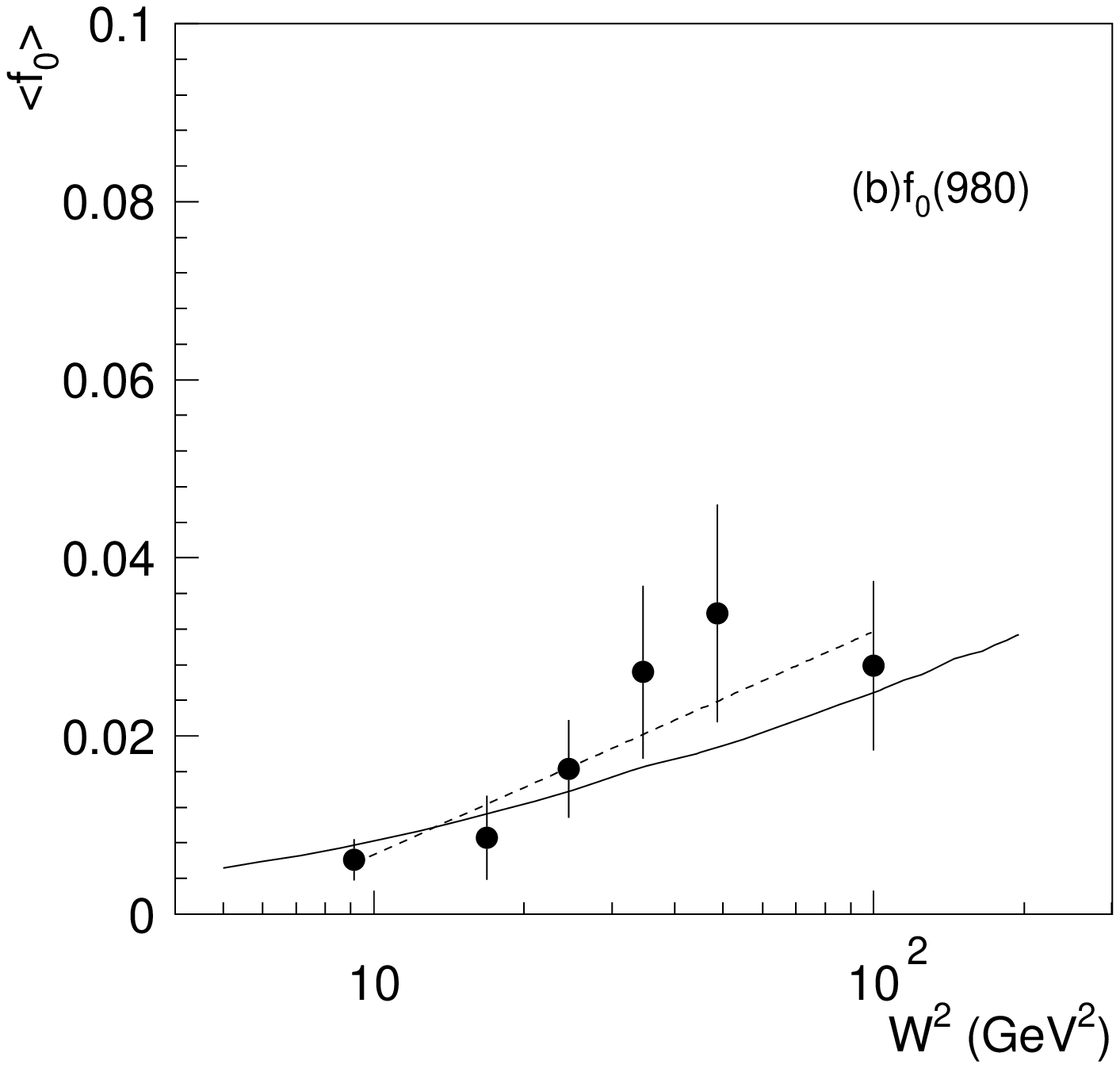}
\end{minipage}%
\begin{minipage}[t]{0.33\linewidth}
\centering
\includegraphics[height=5.2cm]{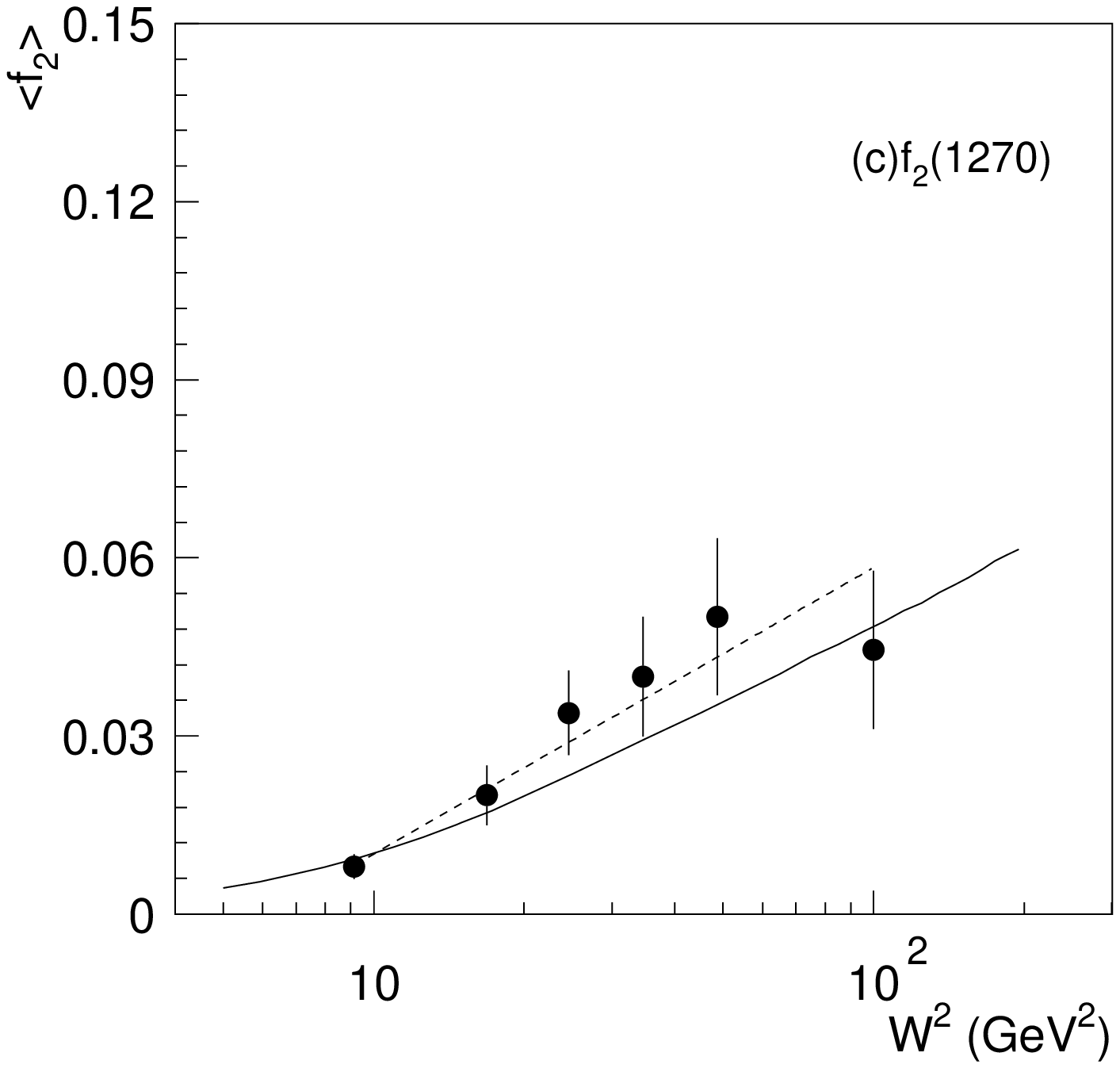}
\end{minipage}%
\caption{\em Average $\rho^{0}(770)$(a), $f_{0}(980)$(b) and $f_{2}(1270)$(c)
multiplicity as a function of $W^{2}$.
The solid line represents the result of the modified Lund simulation.
The dashed line represents a fitted function $a + b \times lnW^{2}$.
The errors are statistical only.}
\label{figure:w2}
\end{figure}

\hspace*{0.5cm}
Fig.~\ref{figure:w2} shows the average multiplicities
as a function of $W^2$. A fit to the average multiplicities with the
expression $a + b \times lnW^{2}$ gives the result presented
in Table~\ref{table:wslope} and shown as a dashed line in Fig.~\ref{figure:w2}.

\vspace{0.3cm}
\begin{table}[hbt]
\setcaptionwidth{10cm}
\begin{center}
\begin{tabular}{|c|c|c|} 
\hline
   Resonance      &       a              &      b            \\
\hline    
 $\rho^{0}(770)$       & $0.026 \pm 0.024$    & $0.062 \pm 0.009$ \\
\hline
 $f_{0}(980)$     & $-0.018 \pm 0.008$   & $0.038 \pm 0.009$ \\
\hline
 $f_{2}(1270)$    & $-0.011 \pm 0.003$   & $0.021 \pm 0.004$ \\
\hline
\end{tabular}
\end{center}
\caption{\em The parameters of the W dependence of the average
multiplicities fitted to the expression $a + b \times lnW^{2}$.
The errors are statistical only.}
\label{table:wslope}
\end{table}  
\vspace{0.3cm}

\begin{figure}[!hb]
\begin{minipage}[t]{0.33\linewidth}
\centering
\includegraphics[height=5.2cm]{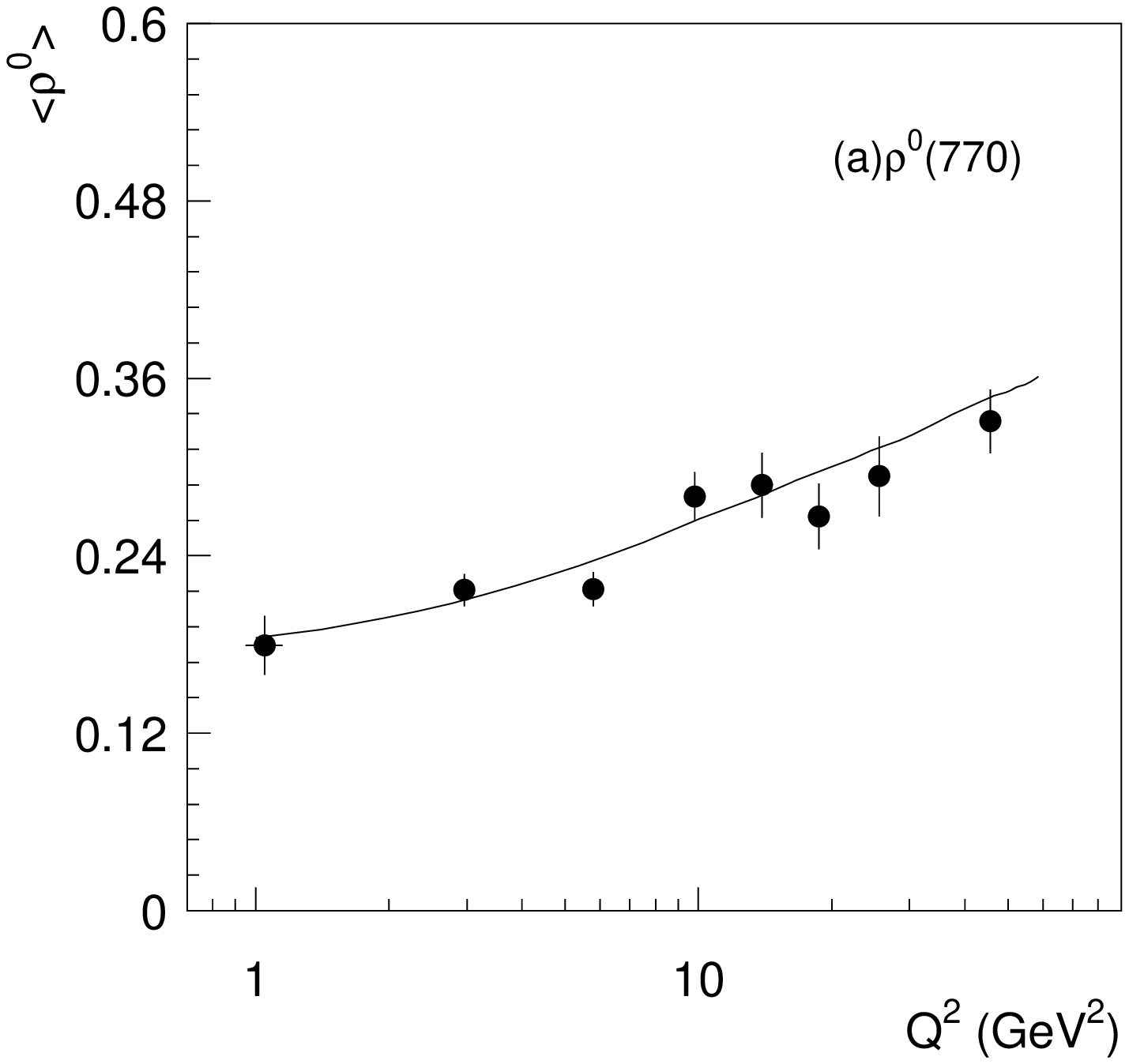}
\end{minipage}%
\begin{minipage}[t]{0.33\linewidth}
\centering
\includegraphics[height=5.2cm]{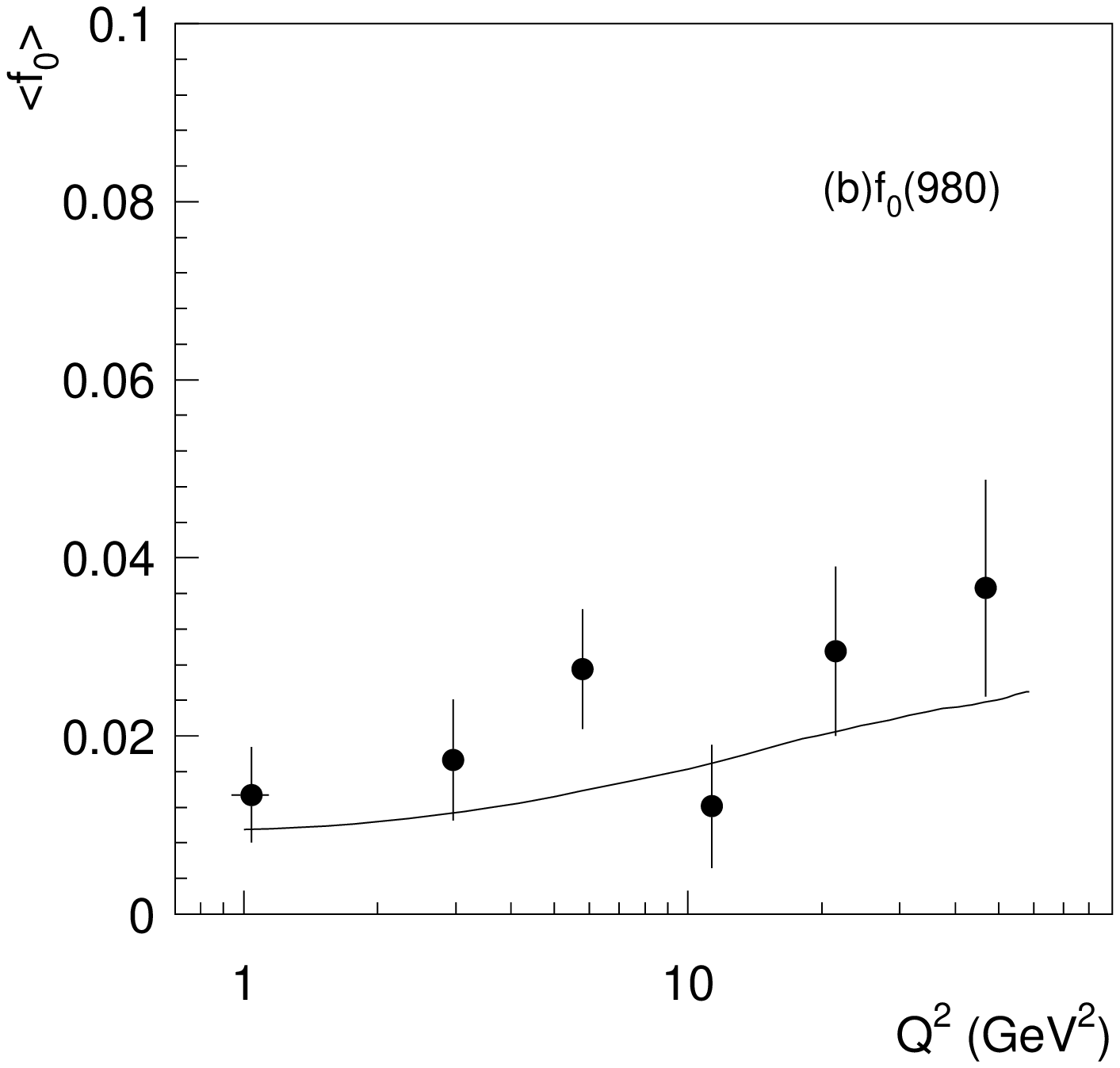}
\end{minipage}%
\begin{minipage}[t]{0.33\linewidth}
\centering
\includegraphics[height=5.2cm]{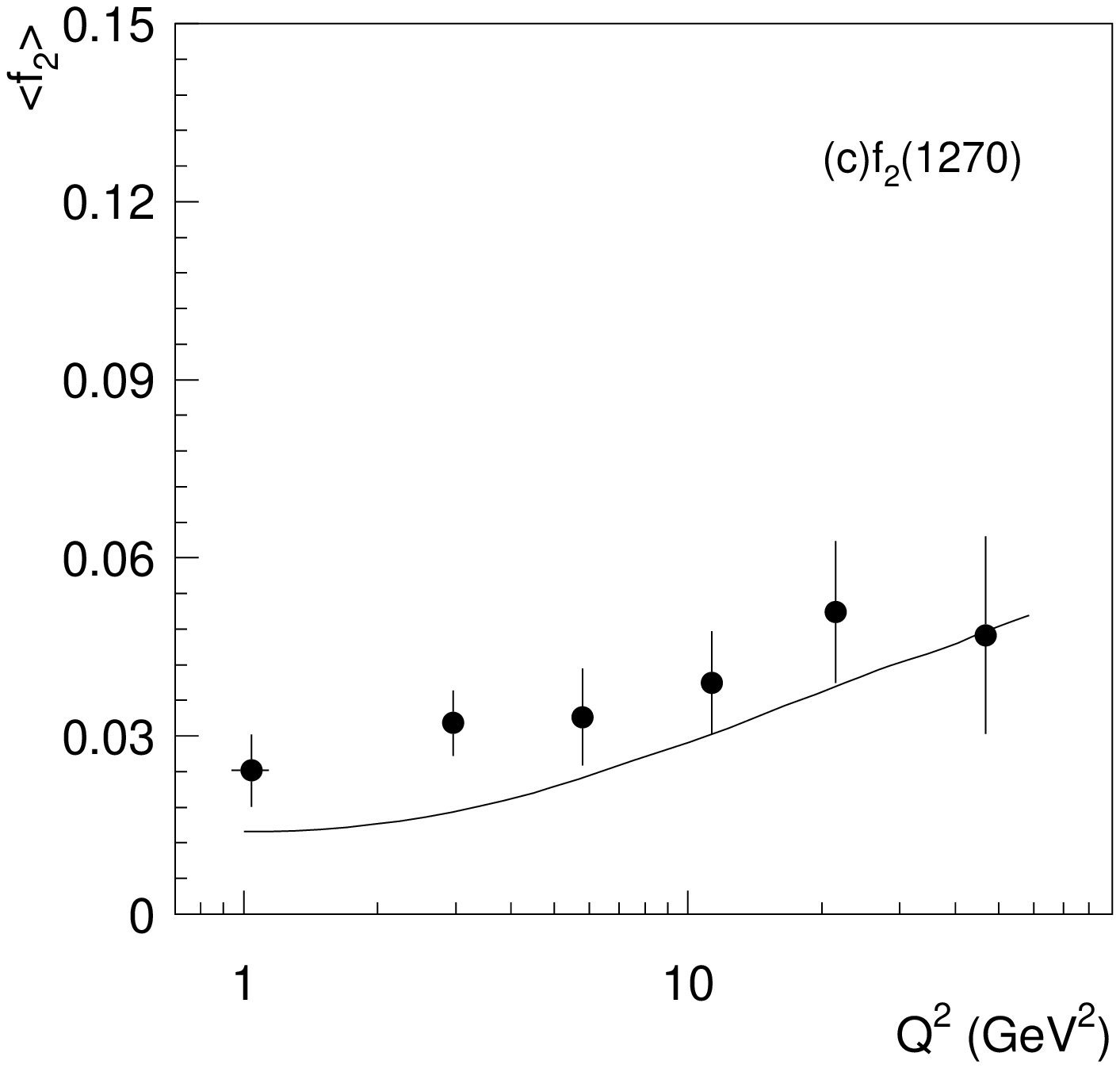}
\end{minipage}%
\caption{\em Average $\rho^{0}(770)$(a), $f_{0}(980)$(b) and $f_{2}(1270)$(c)
multiplicity as a function of $Q^{2}$.
The solid line represents the result of the Lund simulation.}
\label{figure:q2}
\end{figure}

\hspace*{0.5cm}
The dependence of the average multiplicities on $Q^2$
is plotted in Fig.~\ref{figure:q2}. \\

\begin{figure}[!hb]
\begin{minipage}[t]{0.33\linewidth}
\centering
\includegraphics[height=5.0cm]{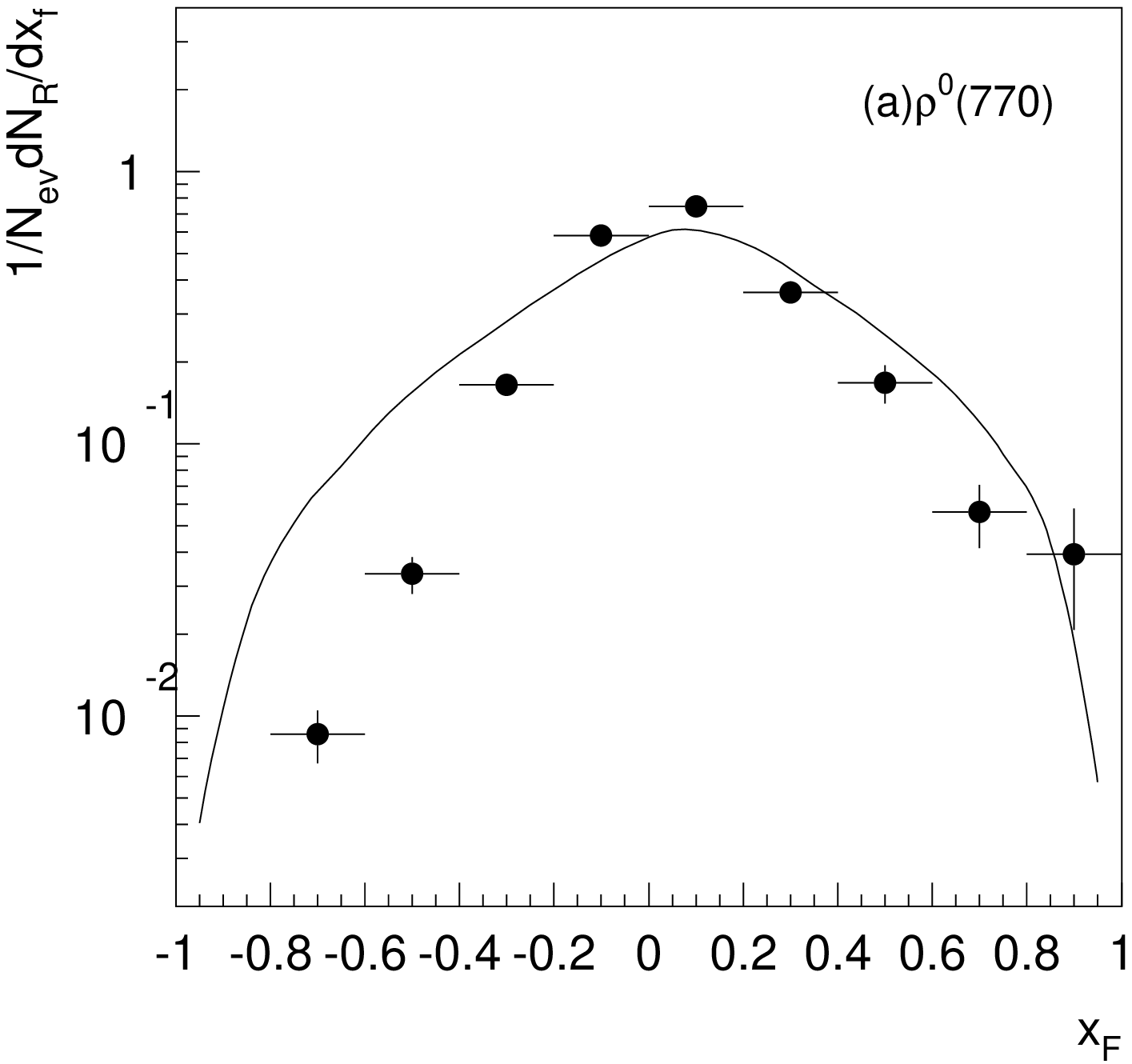}
\end{minipage}%
\begin{minipage}[t]{0.33\linewidth}
\centering
\includegraphics[height=5.0cm]{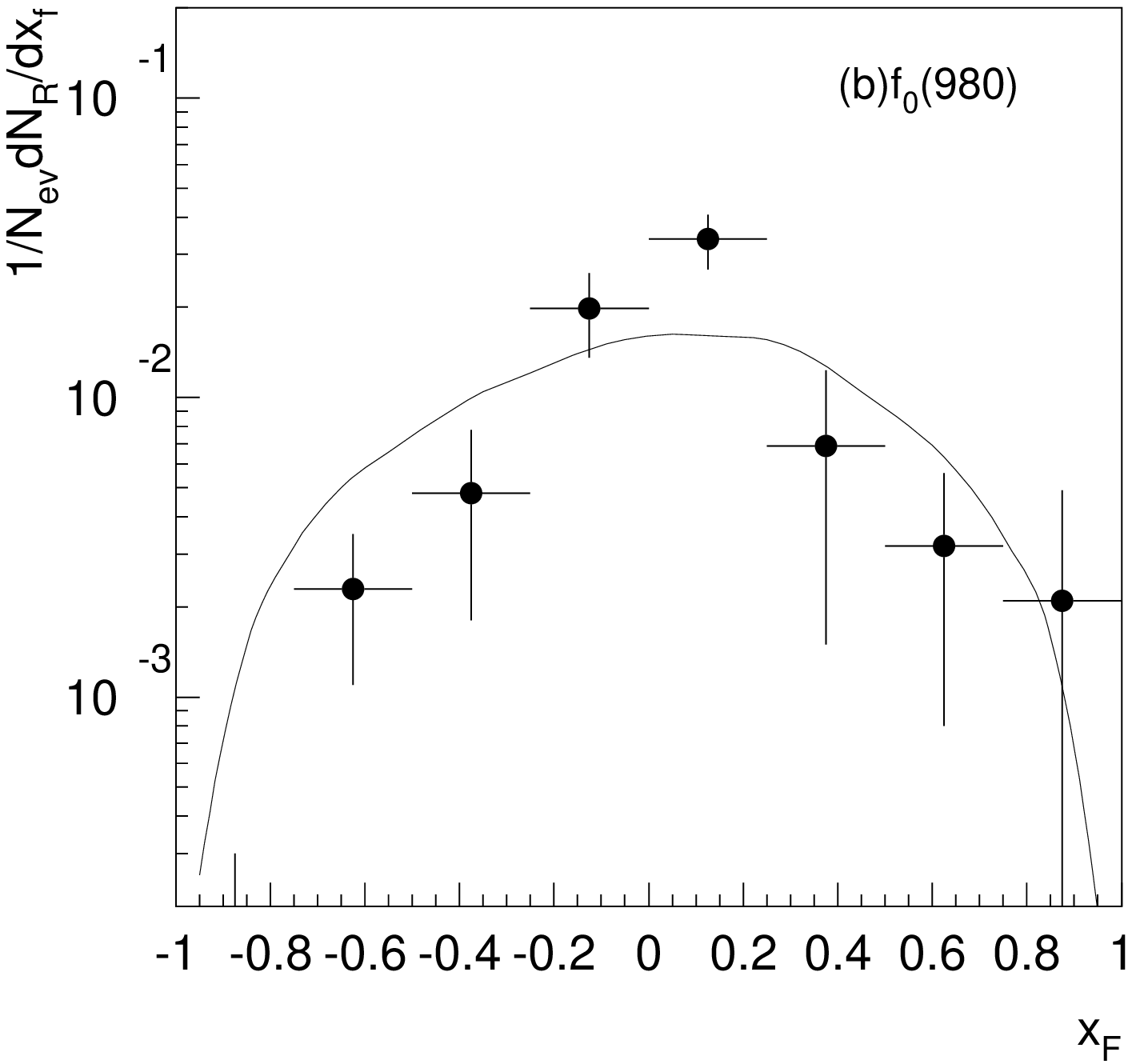}
\end{minipage}%
\begin{minipage}[t]{0.33\linewidth}
\centering
\includegraphics[height=5.0cm]{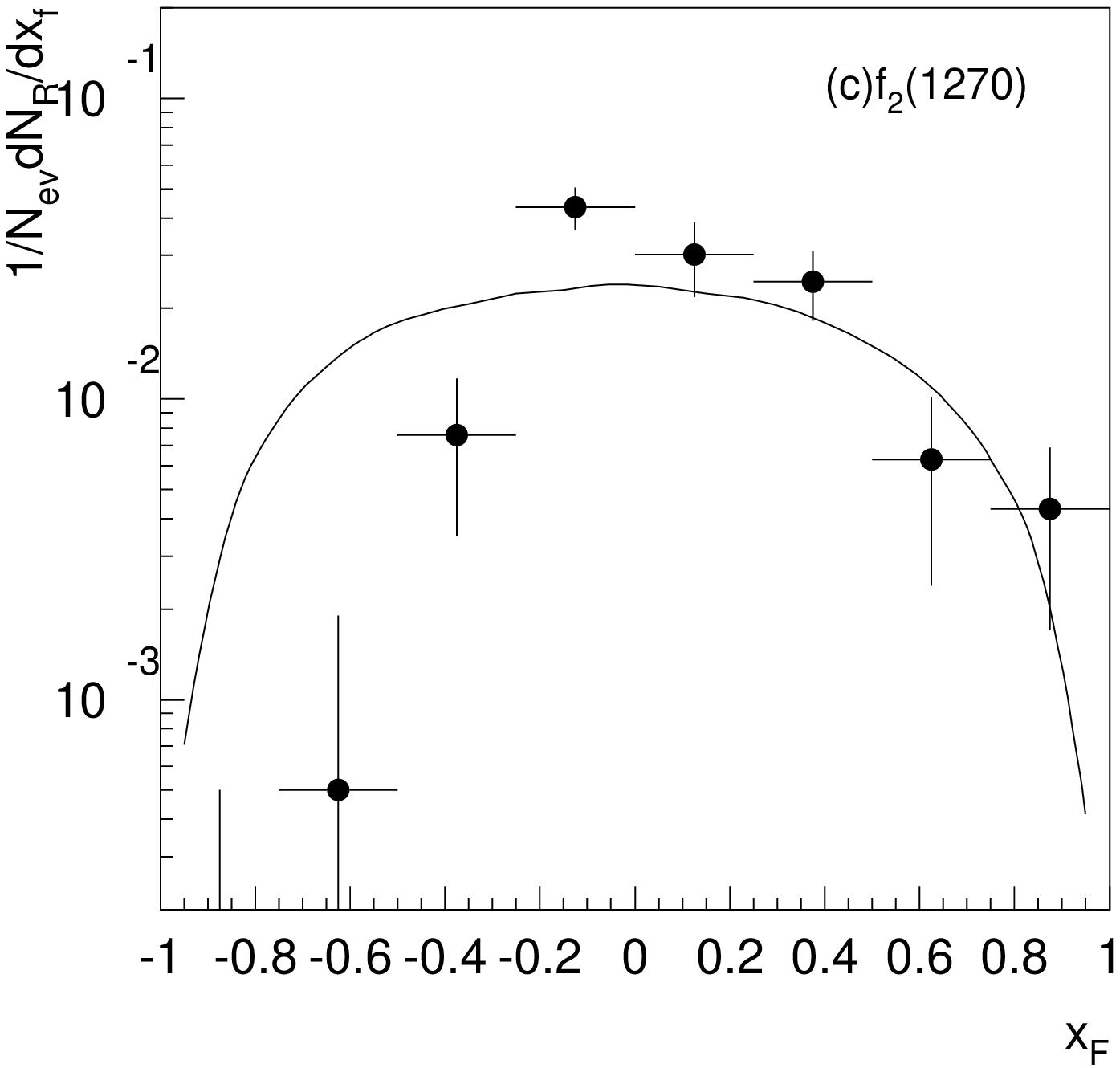}
\end{minipage}%
\caption{\em Average $\rho^{0}(770)$(a), $f_{0}(980)$(b) and $f_{2}(1270)$(c)
multiplicity as a function of $x_F$.
The solid line represents the result of the Lund simulation}
\label{figure:xf}
\end{figure}

\begin{figure}[!hb]
\begin{minipage}[t]{0.33\linewidth}
\centering
\includegraphics[height=5.2cm]{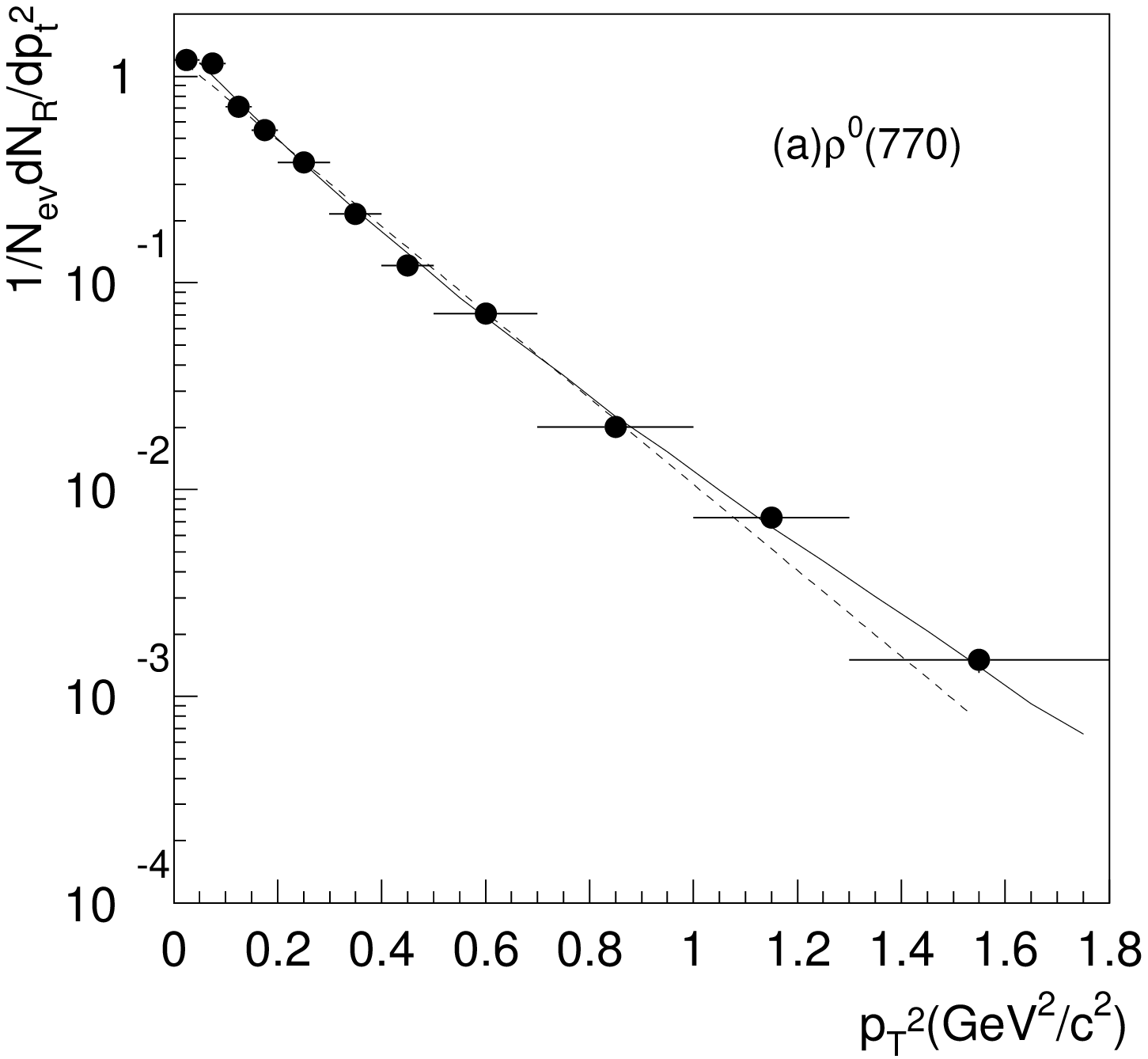}
\end{minipage}%
\begin{minipage}[t]{0.33\linewidth}
\centering
\includegraphics[height=5.2cm]{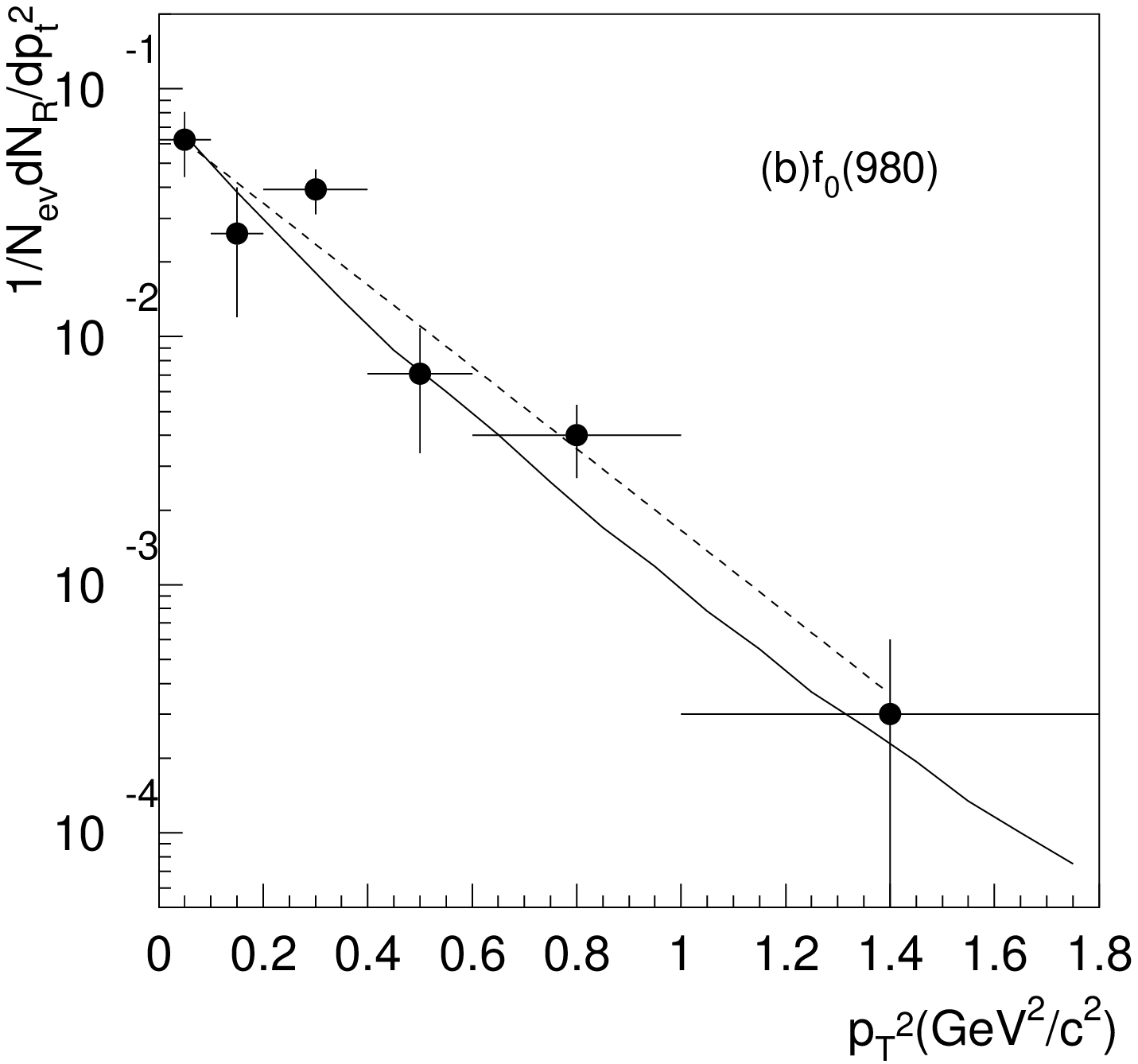}
\end{minipage}%
\begin{minipage}[t]{0.33\linewidth}
\centering
\includegraphics[height=5.2cm]{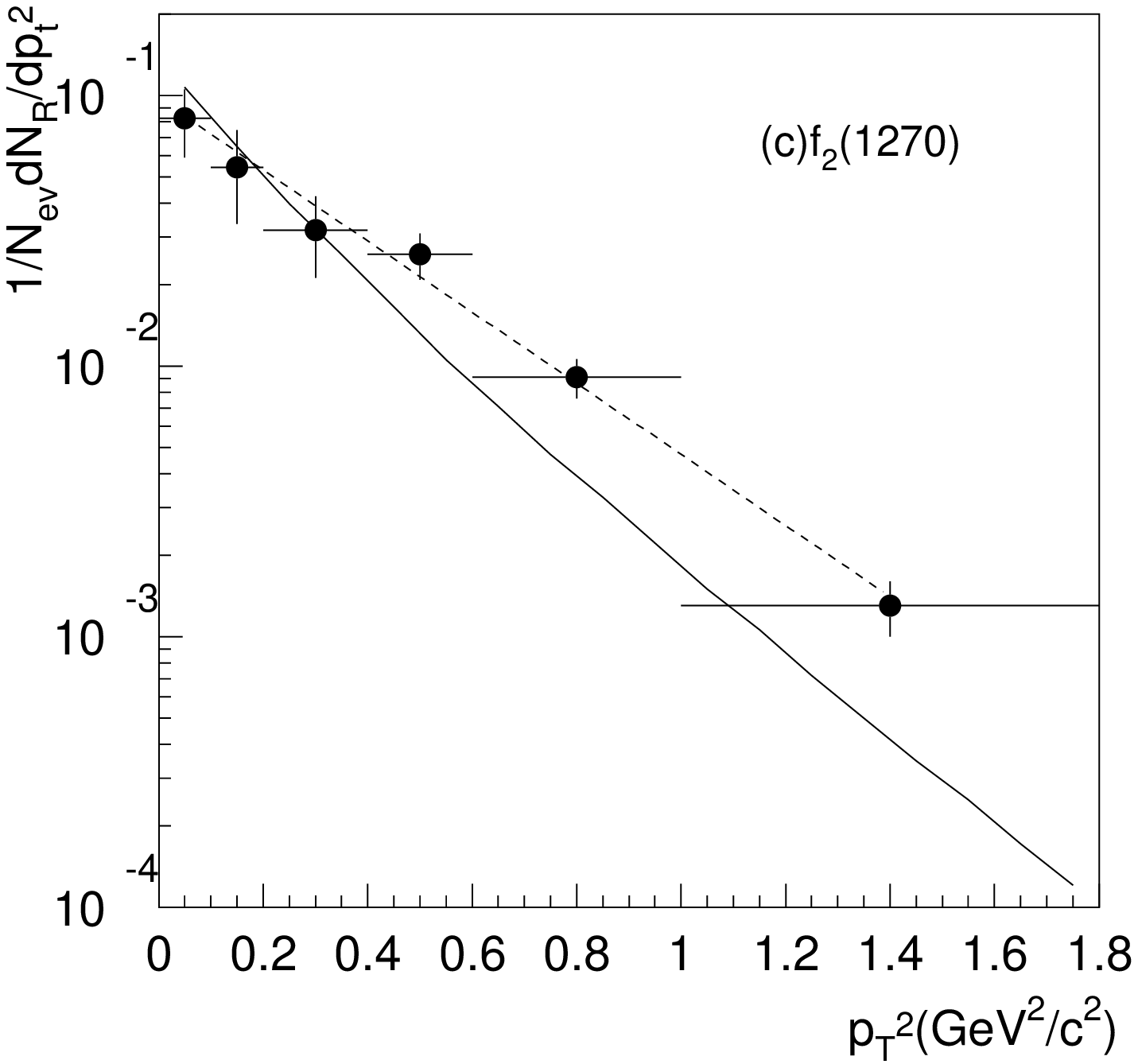}
\end{minipage}%
\caption{\em Average $\rho^{0}(770)$(a), $f_{0}(980)$(b), and $f_{2}(1270)$(c)
multiplicity as a function of $p_{\perp}^{2}$.
The solid line represents the result of the Lund simulation.
The dashed straight line represents the result of the fit by the function
$ae^{-bp_{\perp}^{2}}$.}
\label{figure:pt2}
\end{figure}

\begin{figure}[!hb]
\begin{minipage}[t]{0.33\linewidth}
\centering
\includegraphics[height=5.2cm]{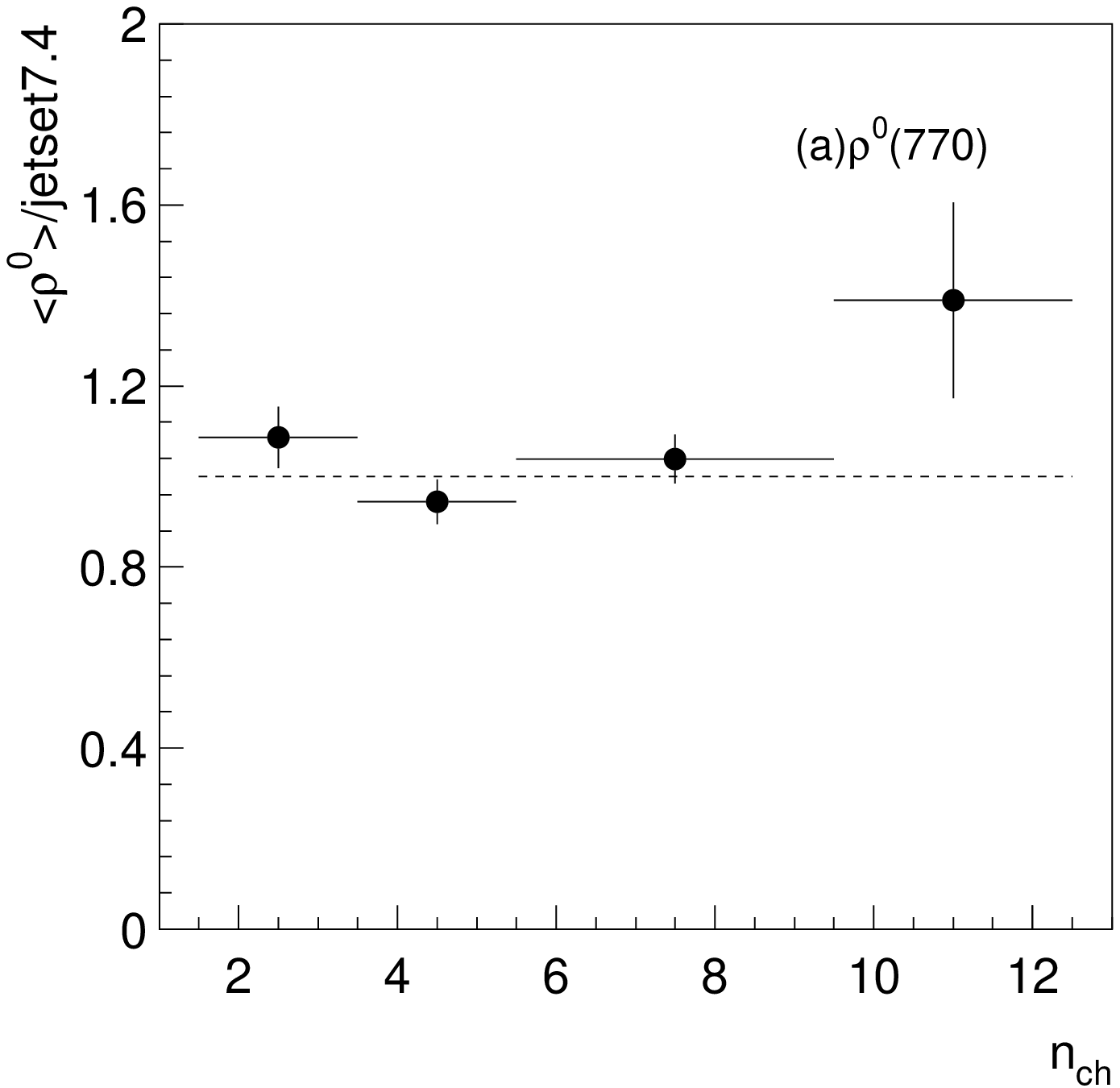}
\end{minipage}%
\begin{minipage}[t]{0.33\linewidth}
\centering
\includegraphics[height=5.2cm]{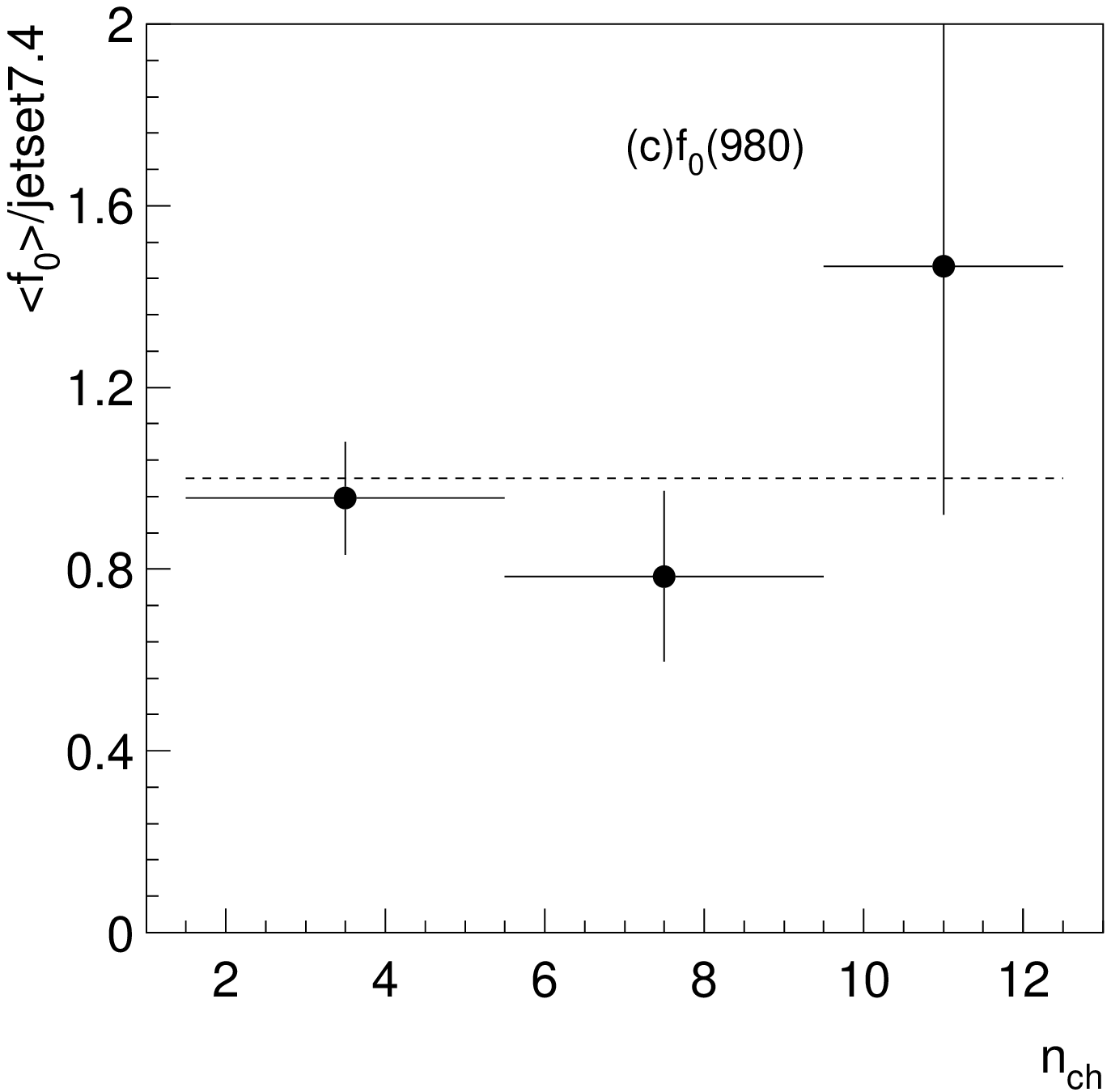}
\end{minipage}%
\begin{minipage}[t]{0.33\linewidth}
\centering
\includegraphics[height=5.2cm]{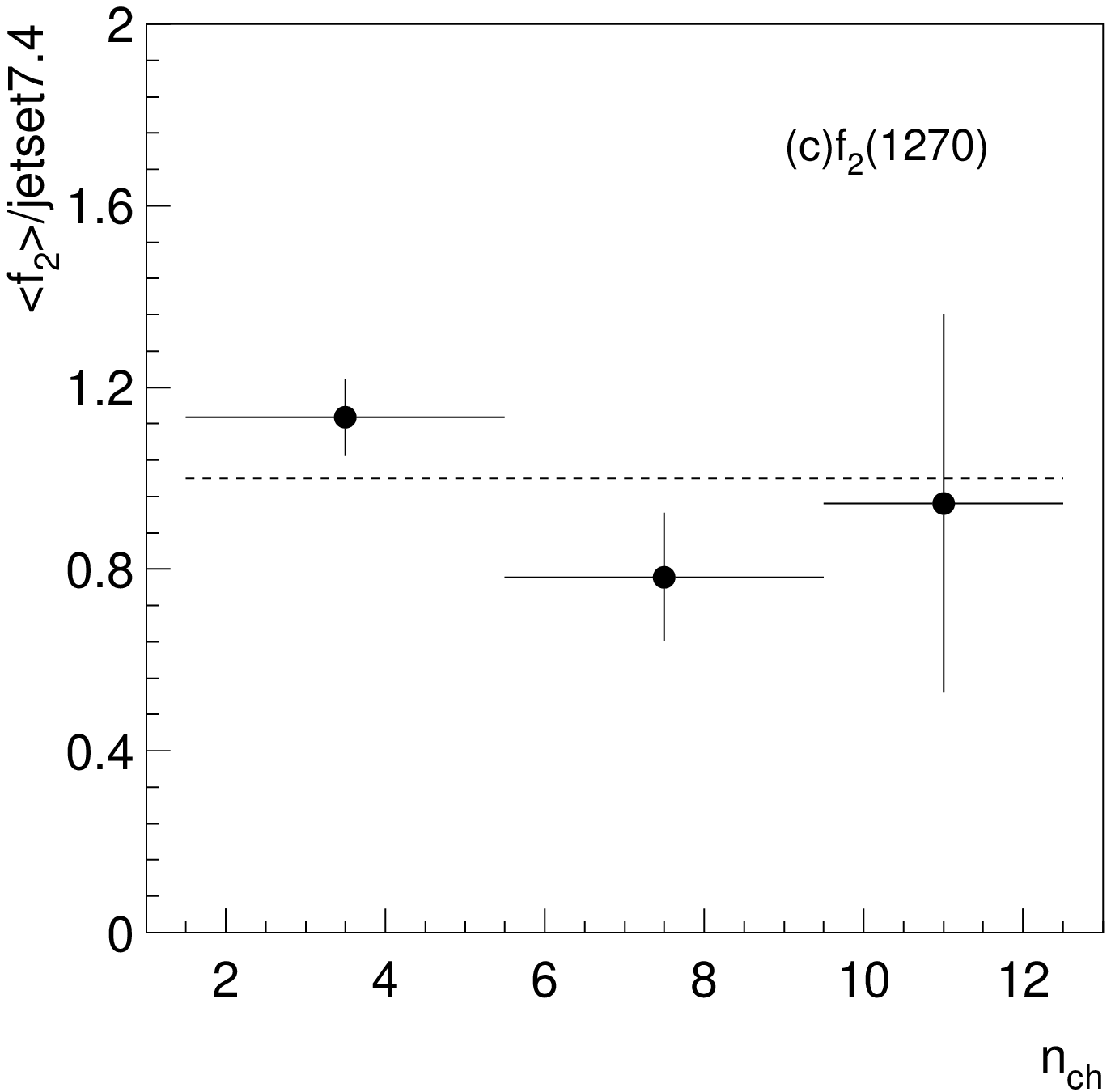}
\end{minipage}%
\caption{\em The resonance multiplicities,
normalized to the ones simulated with the Lund model
as a function of the hadron jet charged particle multiplicity.}
\label{figure:chm}
\end{figure}

\hspace*{0.5cm}
Fig.~\ref{figure:xf} shows the $x_{F}$ dependence of the resonance
production. From this picture we see that the
forward production of $\rho^{0}(770)$ and $f_{2}(1270)$ is enhanced
with respect to the backward production. The Lund model overestimates
the backward production of these resonances. \\
\begin{figure}[!hb]
\begin{minipage}[t]{0.33\linewidth}
\centering
\includegraphics[height=5.0cm]{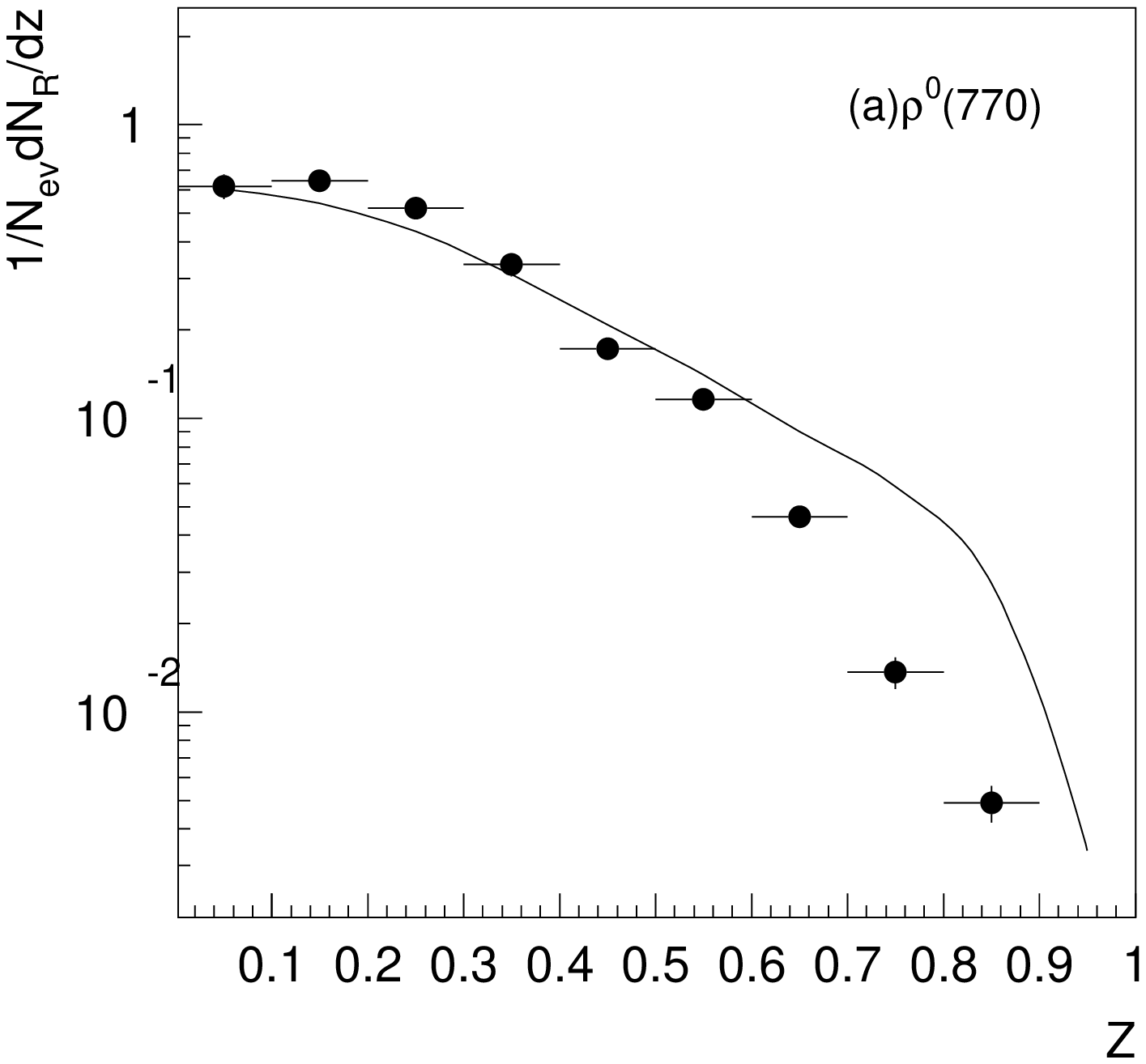}
\end{minipage}%
\begin{minipage}[t]{0.33\linewidth}
\centering
\includegraphics[height=5.0cm]{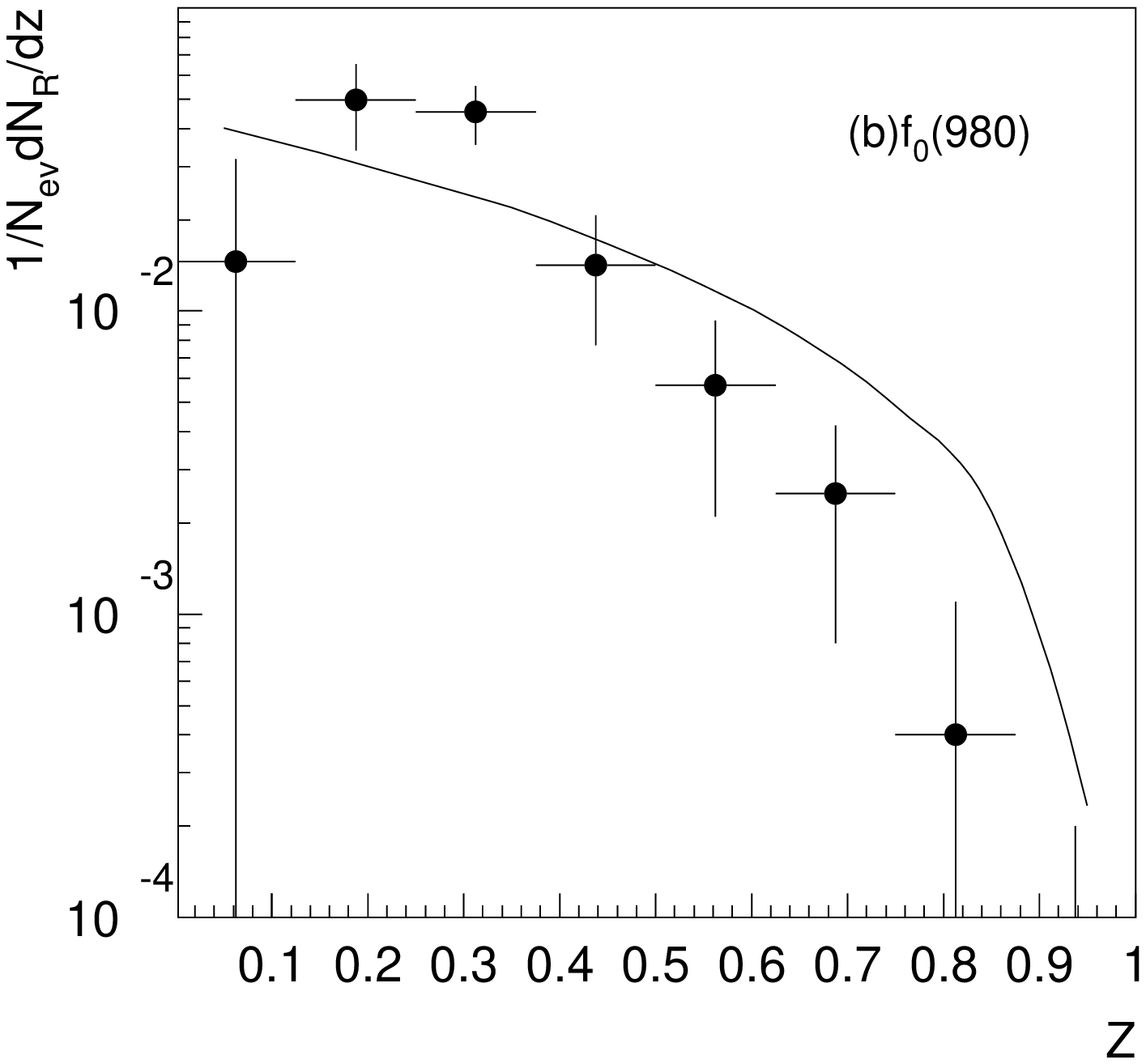}
\end{minipage}%
\begin{minipage}[t]{0.33\linewidth}
\centering
\includegraphics[height=5.0cm]{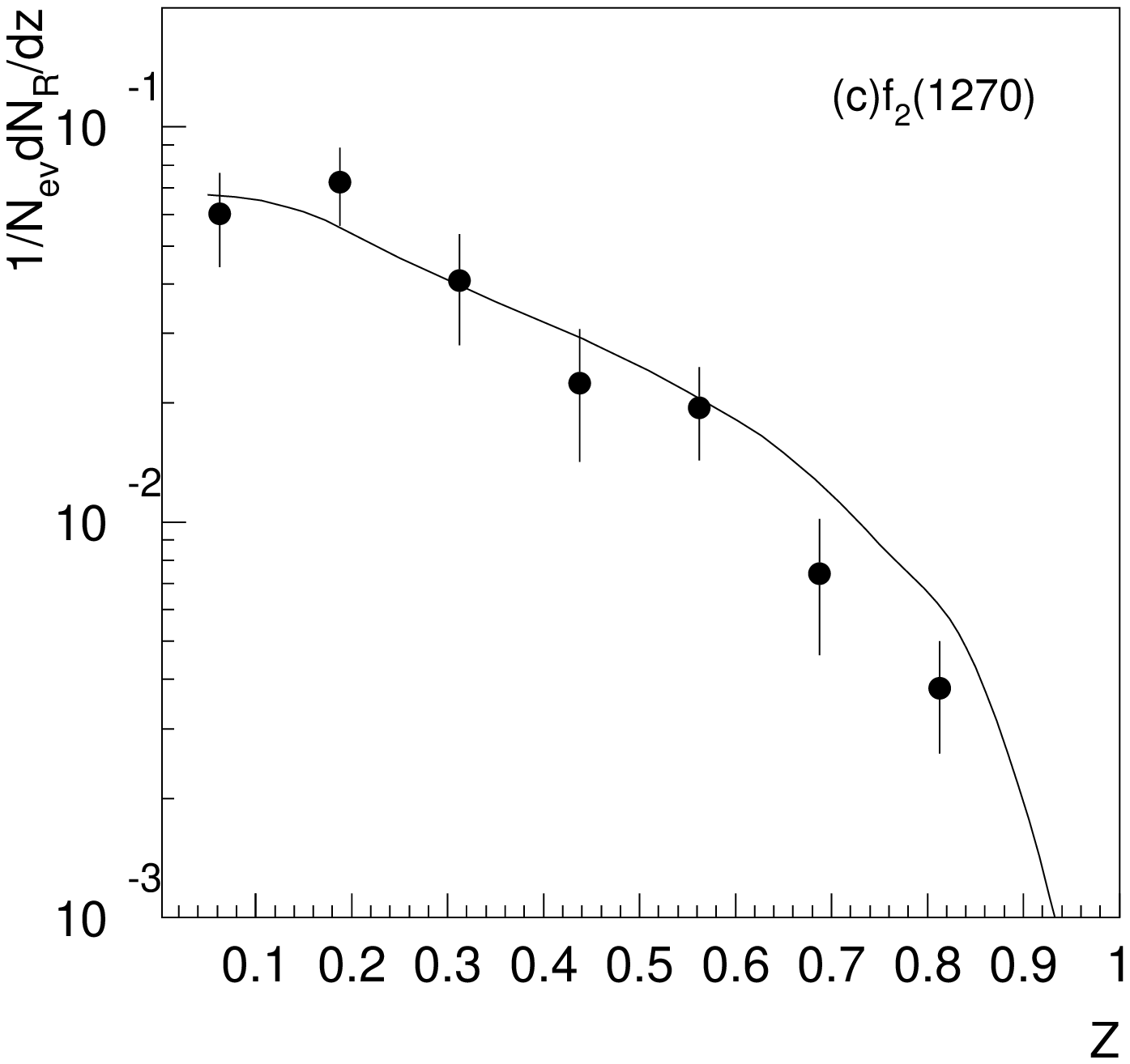}
\end{minipage}%
\caption{\em Average 
$\rho^{0}(770)$(a), $f_{0}(980)$(b) and $f_{2}(1270)$(c) 
multiplicity as a function of $z$ (fragmentation functions).
The solid line represents the result of the Lund simulation.}
\label{figure:zzf}
\end{figure}
\hspace*{0.5cm}
The resonance fragmentation function is shown
in Fig.~\ref{figure:zzf}. The high $z$ region of the distributions is
overestimated by the Lund model. \\ 
\hspace*{0.5cm}
The $p_{\perp}^{2}$ dependence of the resonance production
is shown in Fig.~\ref{figure:pt2}.
In the case of $\rho^{0}(770)$ our result favours a steeper $p_{\perp}^{2}$
slope than previously measured in \cite{nu5,nu7} and agrees with
that of ref. \cite{nu10}.  
A fit using the function $ae^{-bp_{\perp}^{2}}$ gives
\begin{center}
$a=(1.28\pm 0.04)($GeV/c$)^{-2}$,
$b=(4.79\pm 0.07)($GeV/c$)^{-2}$.
\end{center}
\hspace*{0.5cm}
An experiment \cite{nu5} found that the slope parameter $b$
in the range $p_{\perp}^{2} > 0.5($GeV/c$)^{2}$ is different from the slope
below this value. We find
\begin{center}
$b=(5.3\pm 0.2)($GeV/c$)^{-2}$.
\end{center}
for $p_{\perp}^{2} < 0.5($GeV/c$)^{2}$; and
\begin{center}
$b=(4.1\pm 0.1)($GeV/c$)^{-2}$.
\end{center}
for $p_{\perp}^{2} > 0.5($GeV/c$)^{2}$. Thus we confirm the existence
of two different slopes as found earlier \cite{nu5}. \\
\hspace*{0.5cm}
We also observe that the $p_{\perp}^{2}$ distribution for $f_{2}(1270)$
in the data is harder than the results of the Lund model simulation.
This could suggest a mass dependence of the $p_{\perp}^{2}$ behaviour. \\
\hspace*{0.5cm}
Finally, we address the possibility that the $f_{0}(980)$(b)
corresponds to the vacuum scalar state suggested by Gribov
(see Section~\ref{sec:introduction}).
Fig.~\ref{figure:chm} shows the resonance multiplicities 
divided by the ones simulated with the Lund model
as a function of the hadron jet charged track multiplicity. The measured
and MC simulated resonance multiplicities were normalized to the
same average multiplicity.
From Fig.~\ref{figure:chm} we conclude that in our $W$ range
and at our level of accuracy (15 - 20\%) we do not observe
the $f_{0}(980)$ excess at low multiplicities predicted by this model. \\

\section{Systematic uncertainties}
\label{sec:systematics}

\hspace*{0.5cm}
The main source of systematic uncertainty results from the subtraction
of the reflections. Reflections in the data can be
different from the MC not only by a single overall factor but
could also have a different shape due to different relative
resonance yields. We estimated this effect as follows.
The fit to the $\pi^+\pi^-$ invariant mass distribution gave the reflection
normalization factor $p$. We repeated the fit
changing $p$ beyond the statistical uncertainty obtained from the
initial fit (up to 15\%).
This results in relative changes of the
resonances yields as reported in Table~\ref{table:errsys}. \\
\hspace*{0.5cm}
Other sources of systematics are:
\begin{itemize}

\item The uncertainty due to the choice of the fit boundaries: increasing 
      the left boundary by 25 MeV or lowering the right boundary by 300 MeV
      changes the yields as shown in Table~\ref{table:errsys};

\item Removing the cubic or square term in the exponential function 
      used to describe the background changes the result as shown
      in the line labelled "Fit model" of Table~\ref{table:errsys};

\item Changing the bin width has a negligible effect on the result.

\end{itemize}
\vspace{0.2cm}
\begin{table}[hbt]
\setcaptionwidth{10cm}
\begin{center}
\begin{tabular}{|c|c|c|c|} 
\hline
  Source of systematics & $\rho^{0}(770)$ & $f_{0}(980)$ & $f_{2}(1270)$ \\
\hline    
 Reflection subtraction &   8\%    &   10\%      &  9\%  \\
\hline
 Fit boundaries &   4\%   &   8\%      &  9\%  \\
\hline
 Fit model &   2\%    &   9\%      &  13\%  \\
\hline
 Total    &   9\%   &   16\%      &  18\%  \\
\hline
\end{tabular}
\end{center}
\caption{\em Systematic uncertainties on the resonance yields.}
\label{table:errsys}
\end{table}  
\vspace{0.2cm}
\hspace*{0.5cm}
The overall systematic uncertainty on the measured resonance yields, also
shown in Table~\ref{table:errsys}, is obtained
by adding all effects in quadrature. \\
\hspace*{0.5cm}
The systematic uncertainties affecting the resonance widths
are shown in Table~\ref{table:errwidth}.

\begin{table}[hbt]
\setcaptionwidth{10cm}
\begin{center}
\begin{tabular}{|c|c|c|c|} 
\hline
  Source of systematics & $\rho^{0}(770)$ & $f_{0}(980)$ & $f_{2}(1270)$ \\
\hline    
 Reflection subtraction &   13     &  3    & 11   \\
\hline
 Bin width &  5     &  3      & 1   \\
\hline
 Fit model &  12     &  5     &  6   \\
\hline
 Total    &  18    &  7       &  13   \\
\hline
\end{tabular}
\end{center}
\caption{\em Systematic uncertainties on the resonance widths, in MeV.}
\label{table:errwidth}
\end{table}  

\hspace*{0.5cm}
Table~\ref{table:resnumbfin} summarizes our results with statistical
and systematic uncertainties added in quadrature.

\begin{table}[hbt]
\setcaptionwidth{10cm}
\begin{center}
\begin{tabular}{|l|c|c|c|c|} 
\hline
 Resonance      & Number of & Average      &  Mass & $ \Gamma$(MeV)\\
                & resonances    & Multiplicity & (MeV) &  \\ 
\hline    
 $\rho^{0}(770)$ & 130368$\pm$12509 & 0.195$\pm$0.019 & 768$\pm$2.5 &151$\pm$19\\
\hline
 $f_{0}(980)$    &  11809$\pm$2726 & 0.018$\pm$0.004 & 963$\pm$5 & 35$\pm$12\\      
\hline
 $f_{2}(1270)$   &  25189$\pm$6018 & 0.038$\pm$0.009 & 1286$\pm$11 &198$\pm$32\\
\hline
\end{tabular}
\end{center}
\caption{\em Total corrected numbers and yields of resonances and their
masses and widths. The errors include statistical and systematic error.}
\label{table:resnumbfin}
\end{table}

\section{Yields of {\boldmath $\rho^0(770)$} and comparison with previous experiments}
\label{sec:compprev}

The average $\rho^{0}(770)$ multiplicity measured in the NOMAD experiment
for neutrino and antineutrino charged current interactions
is shown in Table~\ref{table:avmul}, together with the results from
previous experiments. 
In the neutrino charged current sample, our measurement confirms
the results of all previous experiments in a similar W range with a much
better statistical accuracy. The deviation of the SKAT \cite{nu11} result 
is usually explained as due to the lower average $W$. 
In the case of $\bar\nu_{\mu}$,
there was some disagreement between previous experiments which could not
be explained by different $W$ ranges.
Our result confirms the measurements in which
the average $\rho^{0}(770)$ multiplicity in $\overline\nu_{\mu}CC$ interactions
is the same as in $\nu_{\mu}CC$ interactions.

\vspace{0.5cm}
\begin{table}[hbt]
\setcaptionwidth{10cm}
\begin{center}
\begin{tabular}{|c|c|l|c|} 
\hline
   Reaction       &  W            & Average multiplicity &    Reference  \\
\hline    
 $\nu_{\mu}$nucleon     &  W$>$2GeV   & 0.195$\pm$0.007$\pm$0.019& this analysis
\\       
 $\nu_{\mu}$p     &    5.5      & 0.18$\pm$0.02$\pm$0.02  & \cite{nu5}   \\        
 $\nu_{\mu}$p     &   W$>$2GeV  & 0.21$\pm$0.03$\pm$0.03  & \cite{nu10}   \\       
 $\nu_{\mu}$p     &    4.9      & 0.16$\pm$0.02  &   \cite{nu8}     \\    
 $\nu_{\mu}$p     &  W$>$2GeV   & 0.20$\pm$0.05  &   \cite{nu15}    \\       
 $\nu_{\mu}$p     &    5.0      & 0.21$\pm$0.04  &   \cite{nu12}    \\    
 $\nu_{\mu}$n     &  W$>$2GeV   & 0.21$\pm$0.02$\pm$0.03  &   \cite{nu10}  \\     
 $\nu_{\mu}$n     &   -         & 0.17$\pm$0.05  &   \cite{nu15}    \\      
 $\nu_{\mu}$Ne    &    4.8      & 0.17$\pm$0.04  &   \cite{nu6}     \\      
 $\nu_{\mu}$Freon &    3.1      & 0.09$\pm$0.02  &   \cite{nu11}    \\  
\hline
 $\overline\nu_{\mu}$nucleon  & W$>$2GeV & 0.18$\pm$0.02$\pm$0.02 & this analysis \\  
 $\overline\nu_{\mu}$p  &    4.4   & 0.12$\pm$0.02$\pm$0.02 & \cite{nu5} \\  
 $\overline\nu_{\mu}$p  & W$>$2GeV & 0.20$\pm$0.02$\pm$0.03 & \cite{nu10} \\  
 $\overline\nu_{\mu}$p  &    4.2   & 0.11$\pm$0.02 &  \cite{nu9}         \\
 $\overline\nu_{\mu}$p  &    3.4   & 0.21$\pm$0.03 &  \cite{nu13}        \\
 $\overline\nu_{\mu}$n  & W$>$2GeV & 0.18$\pm$0.03$\pm$0.03 & \cite{nu10} \\  
 $\overline\nu_{\mu}$Ne &    3.9   & 0.12$\pm$0.02 &  \cite{nu7}         \\
 $\overline\nu_{\mu}$Ne &    4.3   & 0.13$\pm$0.02 &  \cite{nu14}        \\
\hline
\end{tabular}
\end{center}
\caption{\em Average $\rho^{0}(770)$
multiplicity in $\nu_{\mu}$ and $\overline\nu_{\mu}$ experiments.}
\label{table:avmul}
\end{table}  

\newpage
\FloatBarrier
\section{Conclusions}
\label{sec:conclusion}

The inclusive production of the meson resonance $\rho^{0}(770)$ in 
neutrino - nucleon charged current
interactions has been studied with much better statistical accuracy
than in previous experiments. \\
\hspace*{0.5cm}
We also looked for other resonances decaying to $\pi^+\pi^-$ pairs.
For the first time the $f_{0}(980)$ meson is observed
in neutrino interactions. The signal has a significance
of about 5 standard deviations. \\
\hspace*{0.5cm}
We see a clear signal of $f_{2}(1270)$. Its production in neutrino
interactions is reliably established. \\
\hspace*{0.5cm}
A global fit to the $\pi^+\pi^-$ pair mass distribution allows us
to obtain the masses and the widths of the $\rho^0(770)$, $f_{0}(980)$ and 
$f_{2}(1270)$.\\
\hspace*{0.5cm}
The average multiplicities of $\rho^{0}(770)$, $f_{0}(980)$ and $f_{2}(1270)$ 
are measured as a function of  $W^{2}, Q^{2}$ and other kinematic
variables. These results are compared with a simulation
based on a modified Lund model. Good agreement for most of the
distributions is found.\\ 
\hspace*{0.5cm}
In addition, the production of $\rho^0(770)$
in $\bar\nu_\mu$~CC interactions has been studied with lower
statistics. This allowed us to clarify the ambiguous experimental
situation on the production of $\rho^0(770)$ in antineutrino interactions. \\
\hspace*{0.5cm}
We see no evidence of the enhanced $f_{0}(980)$ production in low
multiplicity events predicted by Gribov et al. \cite{th3}, \cite{th4}. \\
\hspace*{0.5cm}
These studies demonstrate the capability of NOMAD to obtain
results which up to now have been the exclusive domain
of bubble-chamber experiments, but with much improved statistics.

\section{Acknowledgements}
\label{sec:acknowledgements}

We gratefully acknowledge  the CERN SPS accelerator and beam-line staff
for the magnificent performance of the neutrino beam.\
The experiment was  supported by  the following
funding agencies:
Australian Research Council (ARC) and Department of Industry, Science, and
Resources (DISR), Australia;
Institut National de Physique Nucl\'eaire et Physique des Particules (IN2P3), 
Commissariat \`a l'Energie Atomique (CEA),  France;
Bundesministerium f\"ur Bildung und Forschung (BMBF, contract 05 6DO52), 
Germany; 
Istituto Nazionale di Fisica Nucleare (INFN), Italy;
Joint Institute for Nuclear Research and 
Institute for Nuclear Research of the Russian Academy of Sciences, Russia; 
Fonds National Suisse de la Recherche Scientifique, Switzerland;
Department of Energy, National Science Foundation (grant PHY-9526278), 
the Sloan and the Cottrell Foundations, USA. \\
\hspace*{0.5cm}
We are grateful to V.A. Uvarov, A.A. Minaenko, V.G. Zaec for
the discussions of models and fit methods.

\end{document}